\documentclass[10pt,twocolumn]{IEEEtran}

\usepackage{amsmath}
\usepackage{amssymb}
\usepackage[dvips]{graphicx}
\usepackage{epsfig}

\newcommand{\G}{\mathbf{G}}

\newcommand{\M}{{\mathbf{M}}}

\newcommand{\E}{\mathbb{E}}
\newcommand{\en}{\mathcal{E}}

\newcommand{\p}{{\text{P}}}

\newcommand{\ind}{\mathbf{1}}

\newcommand{\on}{\text{ON}}
\newcommand{\avg}{\text{avg}}

\newcommand{\tmin}{\text{min}}
\newcommand{\g}{\text{g}}

\newcommand{\tsnr}{{\text{\footnotesize{SNR}}}}

\newcommand{\ssnr}{\text{\scriptsize{SNR}}}

\newtheorem{prop}{Proposition}

\begin{document}

\title{Secure Transmission of Delay-Sensitive Data over Wireless Fading Channels}

%
\author{Mustafa Ozmen and M. Cenk Gursoy
\thanks{The authors are with the Department of Electrical
Engineering and Computer Science, Syracuse University, Syracuse, NY, 13244
(e-mail: mozmen@syr.edu, mcgursoy@syr.edu).}
}


%


\maketitle

\begin{abstract}
In this paper, throughput and energy efficiency of secure wireless transmission of delay sensitive data generated by random sources is studied. A fading broadcast model in which the transmitter sends confidential and common messages to two receivers is considered. It is assumed that the common and confidential data, generated from Markovian sources, is stored in buffers prior to transmission, and the transmitter operates under constraints on buffer/delay violation probability. Under such statistical quality of service (QoS) constraints, effective capacity of time-varying wireless transmissions and effective bandwidth of Markovian sources are employed to determine the throughput. In particular, secrecy capacity is used to describe the service rate of buffers containing confidential messages. Moreover, energy per bit is used as the energy efficiency metric and energy efficiency is studied in the low signal-to-noise (SNR) regime. Specifically, minimum energy per bit required for the reliable communication of common and confidential messages is determined and wideband slope expressions are identified. The impact of buffer/delay constraints, correlation between channels, source characteristics/burstiness, channel knowledge at the transmitter, power allocation, and secrecy requirements on the throughput and energy efficiency of common and confidential message transmissions is identified.
\end{abstract}

\thispagestyle{empty}
\section{Introduction} \label{sec:intro}


Inherent broadcast nature of wireless transmissions results in their susceptibility to eavesdropping, which makes security one of the critical considerations in wireless networks. One way to address the security problem is by exploiting the attributes of the wireless physical layer. From an information-theoretic perspective, Wyner in \cite{wyner} laid the theoretical foundations of physical-layer security by introducing the wiretap channel wherein the eavesdropper receives a degraded version of the signal received by the legitimate user. In this model, secure communication becomes possible without any shared secret key. As a performance metric, secrecy capacity is defined as the supremum of the achievable communication rates from the transmitter to the legitimate user while the wiretapper is kept ignorant of the information being sent. An extension to a more general wiretap channel model is considered in \cite{csiszar} where the secrecy capacity is derived for nondegraded broadcast channels with confidential messages.

In addition to security considerations, energy efficiency in wireless systems has also been studied intensively in order to address energy costs and environmental concerns \cite{survey-Feng}. The importance of energy-efficient operation has further increased with the unprecedented growth both in mobile data traffic and in the number of mobile devices and networks in recent years. Hence, it is imperative to understand the fundamental performance limits in terms of throughput and energy efficiency in order to utilize the energy/power resources that are scarce especially in mobile scenarios.

Moreover, it is also important to note that while establishing secure links and utilizing the limited resources are critical, yet another concern is that certain quality-of-service (QoS) guarantees need to be provided particularly for delay-sensitive data traffic. For instance, in a number of applications such as voice over IP (VoIP), multimedia streaming, interactive video, and online gaming, constraints on delay, packet loss, or buffer overflow probabilities are imposed in the system design so that the end-users experience satisfactory performance levels. In this setting, effective QoS provisioning for delay-sensitive data traffic depends on the accuracy of the source traffic models. For instance, voice traffic can be modeled as an ON/OFF Markov process, and for variable bit-rate video traffic, autoregressive, Markovian, or Markov-modulated processes can be used \cite{survey-VBRvideotraffic}.

With the above-mentioned motivations, our primary goal in this paper is to identify the throughput and energy efficiency of secure wireless transmissions in the presence of statistical QoS requirements of delay-sensitive data traffic generated by random sources.

\subsection{Literature Overview}

As noted above, addressing security considerations is essential in wireless communication networks due to the ease in eavesdropping of wireless transmissions. With this motivation, information-theoretic security has been extensively investigated. For instance, in \cite{liang-broad} and \cite{boundsecrecy} wiretap channels with fading have been studied whereas authors in \cite{mimowiretap1} and \cite{mimowiretap2} incorporated the multiple antenna settings to wiretap channels. Furthermore, the energy efficiency of secure and reliable communication schemes have been addressed in several recent studies.  The work in \cite{securelowsnr} addressed secure communication in the low signal-to-noise ratio (SNR) regime and identified the minimum energy \emph{per secret bit} and the wideband slope (which are two key performance metrics in the low SNR regime \cite{verdu}).  Motivated similarly by energy efficiency requirements, Comaniciu and Poor in \cite{eesecrecy1} investigated the security-energy tradeoff from an information theoretic perspective. Zhang \emph{et al.} in \cite{eesecrecy2} studied three-node MIMO wiretap channels in order to design an energy efficient precoder. Ng \emph{et al.} in \cite{eesecrecy3} considered secure OFDMA systems and addressed the energy efficient resource allocation problem. Kalantari \emph{et al.} in \cite{eesecrecy4} investigated the power control in wiretap interference channels where users either work together or act as selfish nodes. Similar to our motivation, Chen and Lei in \cite{eesecrecy5} took energy efficiency, security and QoS guarantees into account jointly and worked on maximizing the secrecy energy efficiency while having constraints on delay. In \cite{zhu1} and \cite{zhu2}, Zhu  \emph{et al.} investigated the cross layer scheduling of OFDMA networks with both open and private data transmissions. In \cite{fixedsecrecy}, two medium-access protocols were proposed and the mean service rate, the source's data queue and the secret keys queue was analyzed. Shafie and Al-Dhahir \cite{shafie} proposed a network scheme that consists of a source node and a destination in the presence of buffer aided relay node and an eavesdropper, while taking the data burstiness of source and energy recycling process at the relay into account. In \cite{shafie2}, secure and stable throughput region is investigated by employing beamforming based cooperative jamming that depends on the channel side information available at the transmitter. In \cite{mao}, authors assumed that only the distribution of eavesdropper is known at the transmitter and studied the problem that maximizes the long-term data admission rate while having constraints on the secrecy outage and stability of the data queue. Khalil \emph{et al.} in \cite{khalil} derived upper and lower bounds on the secrecy capacity of the flat fading channel with limitations on delay.
For more details regarding the advances in this rich field of physical-layer security in wireless communications, we refer to surveys and overviews provided in \cite{liang-foundtrend}--\cite{Zou}.

As a theory to address the delay and other deterministic service guarantees, network calculus has been introduced by Cruz in early 1990s \cite{cruz_part1}, \cite{cruz_part2}. Thereafter, Chang in \cite{chang} introduced the theory of effective bandwidth of a time-varying source as a stochastic version of the network calculus \cite{Changbook}, \cite{ChangZajic}. Effective bandwidth theory identified the performance and resource requirements in the presence of statistical QoS constraints which are imposed as limitations on buffer/delay violation probabilities. Effective bandwidths of various source models have been studied in the literature. For instance,  Elwalid and Mitra in \cite{elwalid} investigated the effective bandwidth of Markovian traffic sources while Markov fluids, Markov-modulated Poisson sources, and general stationary sources were addressed in \cite{ebw} and \cite{costasweber}.

In addition to time varying source characteristics, the channel characteristics vary with time in wireless communications. Hence, in wireless models, accurate characterization of the throughput rely on the understanding of the queueing system operating with time-varying service and arrival rates. Wu and Negi defined the effective capacity \cite{dapeng} as the maximum constant arrival rate that can be supported by time-varying transmission rates. Effective capacity is essentially described as a dual concept to effective bandwidth by applying the theory of effective bandwidth to a model with a time-varying channel capacity and regarding the channel service process as a random source with negative rate. Effective capacity has been employed in determining the performance of wireless systems under QoS constraints (see e.g., \cite{gursoy-Twireless09}--\cite{itjournal} and references therein). We have analyzed secrecy effective capacity and optimal power control in \cite{deli-secrecy}, and energy efficiency under queueing and secrecy constraints in \cite{asilomar-ozmen}, considering only constant-arrival rates.

\subsection{Contributions}

As noted above, wireless physical-layer security has recently been intensively studied. On the other hand, energy costs of security, the tradeoff between energy efficiency and secrecy, and analysis of secure wireless transmissions in the presence of random arrivals and delay/buffer constraints have been addressed up to a lesser degree. Indeed, to the best of our knowledge, most recent studies addressed average delay and stability requirements in this context, and not considered statistical queueing constraints, such as buffer overflow and delay violation limitations, which are frequently imposed especially in real-time applications.  Motivated by these, we study the secure communication of delay-sensitive data traffic generated from Markovian sources (e.g.,  discrete-time Markov, Markov fluid, discrete-time and continuous-time Markov modulated Poisson sources) and investigate the fundamental performance limits of secure throughput and energy efficiency under statistical buffer/delay violation constraints. In particular, we can list the contributions of this paper as follows:

\begin{itemize}
\item Considering two-state (ON/OFF) Markovian source models, throughput expressions for common and confidential messages in terms of source statistics, effective capacity of wireless transmissions of common and confidential messages, and QoS exponent $\theta$ are provided.

\item Energy efficiency metrics, namely the minimum energy per bit and wideband slope, are identified for discrete-time Markov, Markov fluid, and Markov-modulated Poisson arrival models again in terms of important system, channel, and source parameters.

\item The effect of source randomness, channel correlation, secrecy requirements, buffer/delay QoS constraints on the performance metrics are identified for both common and confidential messages from both analytical characterizations and numerical results.

\item Throughput and energy efficiency metrics are obtained when the transmitter knows the channel statistics but not the realizations of the channel fading, and therefore sends the confidential data at a fixed rate.
\end{itemize}

\section{Channel Model} \label{sec:channelmodel}
\begin{figure}
\begin{center}
\includegraphics[width=0.45\textwidth]{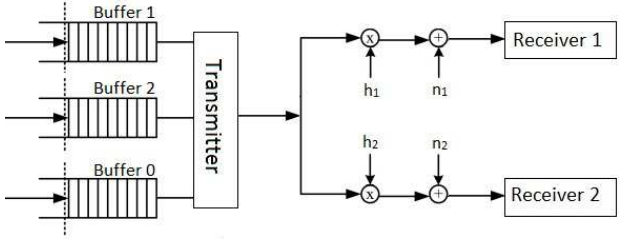}
\caption{Two-receiver broadcast channel model.}\label{fig:chmodel}
\end{center}
\vspace{-0.4cm}
\end{figure}
As depicted in Figure \ref{fig:chmodel}, we consider a fading broadcast channel in which a transmitter sends common and confidential messages to two receivers. Messages are stored in buffers before being transmitted. Specifically, confidential messages intended for receiver 1 and receiver 2 are kept in buffers labeled 1 and 2, respectively, as shown in Fig. \ref{fig:chmodel}, and common messages are stored in buffer 0. Since delay-sensitive data traffic is considered, statistical queueing constraints are imposed in order to limit buffer overflows and delay violations. We assume flat-fading between the transmitter and receivers. The channel input-output relation can be expressed as
\begin{align}
\begin{array}{ll}
		y_j = h_jx + n_j \text{ for } j = 1,2
	\end{array} \label{eq:i-o}
\end{align}
where $x$ is the channel input and $y_j$ is the output at the $j^{\text{th}}$ receiver for $j \in \{1,2\}$. Input signal includes both confidential and common messages. Average transmitted signal energy is $\E\{|x|^2\} = \en$. Moreover, in (\ref{eq:i-o}), $h_i$ denotes the fading coefficient in the channel between the transmitter and receiver $j$.  Finally, $n_j$ denotes the zero-mean, circularly-symmetric, complex Gaussian background noise at receiver $j$ with variance $\E\{|n_j|^2\} = N_{0}$. Hence, the \emph{input signal-to-noise ratio} (SNR) is
\begin{gather}
\tsnr = \frac{\E\{|x|^2\}}{\E\{|n_j|^2\}} = \frac{\en}{N_{0}} \mbox{ } \quad j=1,2. \label{eq:snr}
\end{gather}
While fading coefficients can have arbitrary distributions with finite energies, we assume that block-fading is experienced. Hence, the realizations of the fading coefficients stay fixed for a block of symbols and change independently for the next block.

\section{Preliminaries} \label{sec:prelim}

In this section, we describe the preliminary concepts employed in our subsequent analysis of secure throughput and energy efficiency in the presence of Markov arrivals and statistical queueing constraints.

\subsection{Effective Bandwidth of Markovian Arrivals} \label{subsec:markovarrivals}

Effective bandwidth characterizes the minimum constant transmission rate required to support the given random data arrival process while satisfying statistical queueing constraints on buffer overflows and delay violations. In this paper, we assume that the data to be sent is generated from Markovian sources and, as noted above, is initially stored in a buffer before transmission. Statistical constraints are imposed on the buffer length. In particular, we assume that the buffer overflow probability satisfies
\begin{equation} \label{eq:theta}
\lim_{q \to \infty} \frac{\log \Pr\{Q \ge q\}}{q} = -\theta
\end{equation}
where $Q$ denotes the stationary queue length, and $\theta$ is the decay rate of the tail distribution of the queue length. The above limiting formula implies that for large $q_{\max}$, we have $\Pr\{Q \ge q_{\max}\} \approx e^{-\theta q_{\max}}$. A closer approximation is \cite{dapeng}
\begin{gather}
\Pr\{Q \ge q\} \approx \varsigma e^{-\theta q} \label{eq:overflowprob-rev}
\end{gather}
where $\varsigma = \Pr\{Q > 0\}$ is the probability of non-empty buffer. Hence, for a sufficiently large threshold, the buffer overflow probability decays exponentially with rate controlled by the QoS exponent $\theta$. Note that as $\theta$ increases, stricter queueing or QoS limitations are imposed.

Conversely, for a given buffer threshold $q$ and overflow probability limit $\epsilon = \Pr\{Q \ge q\}$, the desired value of $\theta$ can be determined as
\begin{gather}
\theta = \frac{1}{q}\log_e \frac{\varsigma}{\epsilon}.
\end{gather}

In the given setting, the delay violation probability is also characterized to decay exponentially and is approximated by \cite{Du-Zhang}
\begin{gather}
\Pr\{D \ge d\} \approx \varsigma e^{-\theta a^*(\theta) d}  \label{eq:delayviolation}
\end{gather}
where $D$ is the queueing delay in the buffer at steady state, $d$ is the delay threshold, and $a^*(\theta)$ is the effective bandwidth of the arrival process, described below.

Let $\{a(k), k=1,2,\ldots\}$ be a sequence of nonnegative random variables, describing the random arrival rates. Also let the time-accumulated arrival process be denoted by $A(t) = \sum_{k=1}^{t} a(k)$. Then, the effective bandwidth is given by the asymptotic logarithmic moment generating function of $A(t)$~\cite{chang}, i.e.,
\begin{gather}
a(\theta)= \lim_{ t \rightarrow \infty}\frac{1}{\theta t}\log \E \left\{e^{\theta A(t)}\right\}.
\end{gather}

We consider four types of Markovian sources, namely discrete-time Markov source, Markov fluid source, discrete-time Markov-modulated Poisson source and continuous-time Markov-modulated Poisson source. We mainly concentrate on a simple two-state (ON-OFF) model. For these sources, we briefly describe below the effective bandwidth, which characterizes the minimum constant transmission (or service) rate required to support the given time-varying data arrivals while the buffer overflow probability satisfies (\ref{eq:theta}).

\subsubsection{Discrete Markov Source} \label{subsubsec:discreteMarkov}
Data arrival process from this source is modeled as a discrete-time Markov chain. As noted above, we consider a two-state Markov chain in which $r$ bits arrive (i.e., the arrival rate is $r$ bits/block) in the ON state while there are no arrivals in the OFF state. The transition probability matrix $\mathbf{J}$ for this two-state source is
\begin{align}
\mathbf{J}=\left[
\begin{matrix}
 p_{11} & p_{12} \\
 p_{21} & p_{22}
\end{matrix}\right] \label{eq:probmatrixtwostate}
\end{align}
where $p_{11}$ denotes the probability of staying in the OFF state and $p_{22}$ denotes the probability of staying in the ON state. The probabilities of transitioning from one state to a different one are therefore denoted by $p_{21} = 1 - p_{22}$ and $p_{12} = 1 - p_{11}$. Given the transition probability matrix, the effective bandwidth is formulated as \cite{chang}
\begin{align}
a(\theta, r) = \frac{1}{\theta} \log_e\!\!\left(\!\!\tfrac{p_{11}+p_{22} e^{r\theta}+\sqrt{ (p_{11}+p_{22}e^{r\theta})^2 - 4(p_{11}+p_{22}-1)e^{r\theta} }  }{2}\right). \label{eq:2discreteEBW}
\end{align}

\subsubsection{Markov Fluid Source}
Data arrival process from a Markov fluid  source is modeled as a continuous-time Markov chain with a generating matrix $\G$. The generating matrix for the two-state case is in the form of
\begin{align}
\G=\left[
\begin{matrix}
 -\alpha & \alpha \\
 \beta & -\beta
\end{matrix}\right] \label{eq:generating}
\end{align}
where $\alpha$ and $\beta$ are the transition rates from one state to another. When the arrival rates for the two-state model are $r$ and $0$ and hence we basically have ON and OFF states, the effective bandwidth is given by
\begin{align}
a(\theta)=\frac{1}{2\theta}\left[\theta r -(\alpha+\beta)+\sqrt{(\theta r -(\alpha+\beta))^2+4\alpha\theta r}  \right]. \label{eq:2fluidEBW}
\end{align}

\subsubsection{Discrete-Time Markov-Modulated Poisson Source}
When data arrivals are modeled as a Poisson process with intensity that is determined by a discrete Markov chain, the source is described as a discrete-time Markov modulated-Poisson process (MMPP). Essentially, discrete-time MMPP is similar to discrete Markov processes but with the difference that the instantaneous arrival rate in each Markov state is Poisson distributed rather than being fixed. Hence, MMPP source has more uncertainty or burstiness comparatively. We again assume that the MMPP source has two states (namely ON and OFF) with different Poisson arrival intensities. In particular, when the source is in the ON state, the Poisson intensity is $r$, while the intensity is zero and hence there are no arrivals in the OFF state. For the Markov chain, we use the same transition probability matrix $\mathbf{J}$ in \eqref{eq:probmatrixtwostate}. Under these assumptions, the effective bandwidth is given in \eqref{eq:discMMPPEB} at the top of the next page.
\begin{figure*}
\begin{align}
a^*(\theta, r) = \frac{1}{\theta} \log_e\!\!\left(\!\!\tfrac{p_{11}+p_{22} e^{r(e^\theta-1)}+\sqrt{ (p_{11}+p_{22}e^{r(e^\theta-1)})^2 - 4(p_{11}+p_{22}-1)e^{r(e^\theta-1)} }  }{2}\right) \label{eq:discMMPPEB}
\end{align}
\end{figure*}
\subsubsection{Continuous-Time Markov-Modulated Poisson Source}
In this case, the data arrival rate is again Poisson distributed but with intensity that varies according to a continuous-time Markov chain. We similarly consider a two-state ON-OFF model and assume that the Poisson arrival intensity is $r$ in the ON state whereas there is no arrival in the OFF state. Employing the same generating matrix $\G$ as in \eqref{eq:generating}, the effective bandwidth can be written as
\begin{align}
\hspace{-0.3cm}a(\theta)=&\frac{1}{2\theta}\left[\left(e^\theta-1\right) r -(\alpha+\beta)\right] \nonumber
\\
&+\frac{1}{2\theta}\sqrt{\big[\left(e^\theta-1\right) r -(\alpha+\beta)\big]^2+4\alpha\left(e^\theta-1\right) r} . \label{eq:2MMPPEBW}
\end{align}

\subsection{Effective Capacity of Wireless Transmissions}

Effective capacity provides the
maximum constant arrival rate that a given time-varying service process can support
while the buffer overflow probability decays exponentially as described in \eqref{eq:theta} \cite{dapeng}. Let $\{R[k], k=1,2,\ldots\}$ denote the discrete-time stationary and
ergodic stochastic service process and $S[t]\triangleq
\sum_{k=1}^{t}R[k]$ be the time-accumulated process. Then, the effective capacity  is given by ~\cite{dapeng}
\begin{equation}
C_E(\tsnr,\theta)=-\lim_{t\rightarrow\infty}\frac{1}{\theta
t}\log_e{\mathbb{E}\{e^{-\theta S[t]}\}}.
\end{equation}
We assume that the fading coefficients $\{h_i\}$ change independently from one block to another. Under this assumption,
the effective capacity simplifies to
\begin{equation}\label{ec}
C_{E}(\tsnr,\theta)=-\frac{1}{\theta}\log_e\mathbb{E}\{e^{-\theta
R}\},
\end{equation}
where $R$ is the instantaneous service (or equivalently data transmission) rate. For instance, the maximum service rate in a single-user fading Gaussian channel is given by the instantaneous channel capacity expressed as
\begin{gather}
R = \log_2(1 + \tsnr z)
\end{gather}
where $z = |h|^2$ denotes the fading power and $h$ is the fading coefficient of the channel. The service rates for the transmission of common and confidential messages in the fading broadcast channel addressed in this paper are described below in Section \ref{subsec:secrecy}.

\subsection{Throughput and Energy Efficiency Metrics} \label{subsec:metrics}

In this section, we formulate the throughput and energy efficiency metrics for wireless links in the presence of random source arrivals, statistical queueing constraints, and time-varying transmission rates. Specifically, we consider two-state Markovian arrival models (described in Section \ref{subsec:markovarrivals}) in which the average arrival rates are $r$ and $0$ in the ON and OFF states, respectively\footnote{For discrete Markov and Markov fluid sources, we have a constant arrival rate of $r$ in the ON state, while the arrival rate is Poisson distributed with intensity (or equivalently average value) $r$ for MMPP sources.}. Stationary distribution of the Markov chains is denoted by $\boldsymbol \pi = [\pi_1, \pi_2]$ where $\pi_1$ and $\pi_2$ are the probabilities of the OFF and ON states, respectively. Therefore, the source average arrival rate is simply
\begin{gather}
r_{\avg}  =\pi_2r = P_{\on}r  \label{eq:avgarrival}
\end{gather}
which is equal to the average departure rate when the queue is in steady state \cite{ChangZajic}. Then, we seek to determine the maximum average arrival rate $r_\avg^*$ that can be supported by the fading channel described in Section \ref{sec:channelmodel} while satisfying the statistical QoS requirements given in the form in (\ref{eq:theta}). As shown in \cite[Theorem 2.1]{ChangZajic}, if the effective bandwidth of the arrival process is equal to the effective capacity of the service process, i.e.,
\begin{align}
a(\theta, r) = C_E(\tsnr, \theta), \label{eq:equalityforQoS}
\end{align}
then, (\ref{eq:theta}) is satisfied, i.e., buffer overflow probability decays exponentially fast with rate controlled by the QoS exponent $\theta$. Hence, we can determine from (\ref{eq:equalityforQoS}) the ON-state maximum arrival rate $r^*(\tsnr, \theta)$ that can be supported by the wireless channel for given $\tsnr$ and QoS exponent $\theta$. Then, the maximum average arrival rate (and hence the throughput) is
\begin{gather}
r_\avg^*(\tsnr, \theta) = r^*(\tsnr, \theta) P_{\on}. \label{eq:avgarrivalrate}
\end{gather}

In this paper, we employ energy per bit as the performance metric of energy efficiency. In our setup, we define energy per bit as
\begin{gather}
\frac{E_b}{N_0} = \frac{\tsnr}{ r_\avg^*(\tsnr,\theta)}.
\end{gather}

In our analysis, following the approach in \cite{verdu}, we study the minimum energy per bit and the wideband slope, which is defined as the
slope of the spectral efficiency curve at zero spectral efficiency.

The minimum energy per bit $\frac{E_b}{N_0}_{\tmin}$ under QoS
constraints can be obtained from \cite{verdu}
\begin{equation}\label{eq:ebnomin-ra}
\frac{E_b}{N_0}_{\tmin}=\lim_{\tsnr\rightarrow0}\frac{\tsnr}{r_\avg^*(\tsnr,\theta)}=\frac{1}{ \dot{r}_\avg^*(0) }.
\end{equation}
At $\frac{E_b}{N_0}_{\tmin}$, the slope $\mathcal {S}_0$ of the
spectral efficiency versus $E_b/N_0$ (in dB) curve
can be found from \cite{verdu}
\begin{equation}\label{eq:widebandslope-ra}
\mathcal{S}_0=-\frac{2\big(\dot{r}_\avg^*(0)\big)^2} { \ddot{r}_\avg^*(0)}\log_e{2}
\end{equation}
where $\dot{r}_\avg^*(0)$ and $\ddot{r}_\avg^*(0)$ are the first and second
derivatives, respectively, of the function $r_\avg^*(\tsnr,\theta)$ with respect to $\tsnr$ at $\tsnr = 0$. $\frac{E_b}{N_0}_{\tmin}$
and $\mathcal{S}_0$ provide a linear approximation of the spectral
efficiency curve at low spectral efficiencies.

\subsection{Instantaneous Secrecy Capacity of Confidential Messages and Capacity of Common Message Transmissions} \label{subsec:secrecy}
In this section, we describe the secrecy capacity in detail in a general case in which the transmitter sends both common and confidential messages\footnote{Here, we consider standard information-theoretic arguments regarding the definition of messages and how they are encoded and transmitted over fading channels (see e.g., \cite{Gopala}, \cite{liang-broad}).} to two receivers, and, with that, we identify the service rates of our queueing model. Confidential and common messages are sent simultaneously and it is assumed that common message is decoded at the receiver in the presence of the interference from the confidential message transmission. Confidential messages of two receivers are sent necessarily using time-division duplexing depending on the channel strengths. More specifically, confidential message is only sent to the receiver with the higher received SNR.

Secrecy capacity quantifies the maximum achievable rates of secure communication. For instance, it is well-known that the secrecy capacity of confidential message transmission with the signal-to-noise ratio denoted by SNR in the presence of an eavesdropper is given by
\begin{align}
R(\tsnr) = \left[\log_2(1 + \tsnr z_m)-\log_2(1 + \tsnr z_e)\right]^+ \label{eq:secrecyrate}.
\end{align}
Note that the above formula of secrecy capacity is a generic one with $z_m$ and $z_e$ denoting the magnitude squares of the fading coefficients of channels of the intended user and eavesdropper, respectively. When the transmitter sends separate confidential messages to each user as we have assumed and described in Section \ref{sec:channelmodel}, the unintended user can be regarded as an eavesdropper.

Having two confidential messages and one common message to send, transmitter allocates its power for the transmission of these messages. We assume that when confidential message intended for receiver $i$ is being sent, $\delta_i$ portion of the power is used for confidential message transmission while $(1-\delta_i)$ portion of the power is used for common message transmission. Additionally, we define the regions for time-division duplexing of confidential messages as
\begin{align}
\Gamma_1&=\left\{(z_1, z_2) \in \mathbb{R}^{2+} : z_1 \geq z_2\right\}, \nonumber
\\
\Gamma_2&=\left\{(z_1, z_2) \in \mathbb{R}^{2+} : z_1 < z_2\right\}. \nonumber
\end{align}
For instance, when we have $(z_1, z_2) \in \Gamma_1$, only confidential message intended for receiver $1$ is transmitted along with the common message\footnote{We note that the event of $z_1=z_2$ occurs with zero probability if the fading powers $z_1$ and $z_2$ have continuous distributions, as frequently assumed in the statistical modeling of the wireless fading channel in the literature. However, in the case of discrete fading distributions, this event is in general a non-zero probability event. In such a case, the  secrecy capacity is zero, and hence no confidential message transmission can be performed. All the power can be allocated to the transmission of the common message by setting $\delta_1=0$.}.
As previously stated, the common message is decoded in the presence of interference from confidential message transmissions. Both users can decode the common message when it is sent at a rate they both can decode, implying that the common message is sent at the minimum rate that both channels can support. Hence, the instantaneous transmission rate of the common message becomes
\begin{align}
R_0(\tsnr)=& \log_2\left(1+\frac{(1-\delta_1)\tsnr z_2}{1+\delta_1\tsnr z_2}\right) \ind\left\{\Gamma_1\right\} \nonumber
\\
&+\log_2\left(1+\frac{(1-\delta_2)\tsnr z_1}{1+\delta_2\tsnr z_1}\right) \ind\left\{\Gamma_2\right\}. \label{eq:rate0}
\end{align}
After subtracting the common message from the received signal, the receiver with the better channel can decode its confidential message without any interference from the common message. Therefore, we can express the instantaneous transmission rate of confidential messages intended for receivers 1 and 2, respectively, as
\begin{align}
R_1(\tsnr)= \log_2\left(\frac{1+\delta_1\tsnr z_1}{1+\delta_1\tsnr z_2}\right) \ind\left\{\Gamma_1\right\} \label{eq:rate1}
\end{align}
\begin{align}
R_2(\tsnr)= \log_2\left(\frac{1+\delta_2\tsnr z_2}{1+\delta_2\tsnr z_1}\right) \ind\left\{\Gamma_2\right\} \label{eq:rate2}
\end{align}
where $\ind\{\cdot\}$ denotes the indicator function\footnote{The secrecy rate expressions in (\ref{eq:rate1}) and (\ref{eq:rate2}) are derived from the generic expression in (\ref{eq:secrecyrate}) For instance, in (\ref{eq:rate1}), $z_1$ and $z_2$ correspond to $z_m$ and $z_e$, respectively, and the signal-to-noise ratio is $\delta_1 \tsnr$. Additionally, the indicator function essentially represents the operation $[\cdot]^+$, ensuring that the secrecy rate is zero if $z_1 < z_2$.}.


\section{Throughput of Secure Transmissions with Random Data Arrivals Under QoS Constraints}

In this section, we investigate the throughput of the transmission of confidential and common messages, considering different random source types introduced in Section \ref{subsec:markovarrivals}. In order to highlight the impact of random arrivals, we also address the case of a source with a constant arrival rate. For each source type, we characterize the maximum \emph{average} arrival rate as the maximum throughput. Thus, we determine the throughput by deriving the maximum average arrival rate in terms of SNR for both constant-rate arrivals and the four Markovian arrival models.

We note that our initial analysis considers perfect channel side information (CSI) at the transmitter. Hence, we assume that the transmitter knows the realizations of $z_1$ and $z_2$. This is an accurate assumption, for instance, in a cellular scenario in which the base station knows the channel conditions and the users are not malicious but still the confidential messages are to be kept private from the unintended user. We address the case of no CSI subsequently in Section \ref{section:noCSI}.

\subsubsection{Constant-Rate Source} \label{subsubsec:dMarkov}
 Throughput in the case of constant-rate arrival is given by the    effective capacity. For each message, the effective capacity is given by
\begin{align}
C_{Ei}(\tsnr,\theta_i)=-\frac{1}{\theta_i}\log_e\mathbb{E}\{e^{-\theta_i
R_i(\tsnr)}\} \text{ for } i=0,1,2. \label{eq:sc}
\end{align}
Note that for $i = 1 \text{ and } 2$, we have the maximum constant arrival rates of the confidential messages at the transmitter, which are intended for receivers 1 and 2, respectively. For $i = 0$, we have the maximum constant arrival rate of the common message at the transmitter. Note further that the QoS constraint $\theta_i$ of different messages can in general be different. We also define the function $\g_i(\tsnr)$ as
\begin{align}
\g_i(\tsnr) = \E\left\{e^{-\theta_i R_i(\tsnr)}\right\} = e^{-\theta_i C_{Ei}(\tsnr,\theta_i)}. \label{eq:gdef}
\end{align}
Note that with this definition, we have
\begin{align}
C_{Ei}(\tsnr,\theta_i) = -\frac{1}{\theta_i}\log_e \g_i(\tsnr).
\end{align}
As it will be seen in subsequent subsections, maximum average arrival rates for random sources can also be concisely expressed using the function $\g_i(\tsnr)$.

\subsubsection{Discrete Markov Source} \label{subsubsec:dMarkov}
In this case, we assume that (confidential and/or common) message arrivals to the buffers at the transmitter are according to a discrete-time Markov chain. In the case of ON-OFF discrete Markov source, introducing effective bandwidth expression in (\ref{eq:2discreteEBW}) into \eqref{eq:equalityforQoS},
and solving for $r$, we can obtain the maximum arrival rate $r^*(\tsnr, \theta)$ and then express the maximum average arrival rate as a function of the effective capacity $C_E$ as
\begin{align}
r_{\avg i}^*(\tsnr, \theta_i)\!&=\!\frac{P_{\on}}{\theta_i}\!\log_e\!\left(\!\frac{e^{2\theta_i C_{Ei}(\ssnr, \theta_i)}\! -\! p_{11}e^{\theta_i C_{Ei}(\ssnr, \theta_i)}} {1\!-\!p_{11}\!-\!p_{22}\!+\!p_{22}e^{\theta_i C_{Ei}(\ssnr, \theta_i)}}\!\right)\! \nonumber
\\
\!&=\!\frac{P_\on}{\theta_i}\!\log_e\!\left(\!\frac{1 - p_{11}\g_i(\ssnr) } {(1\!-\!p_{11}\!-\!p_{22})\g_i^2(\ssnr)\! +\!p_{22}\g_i(\ssnr) }\!\right) \label{eq:2discreteravg}
\end{align}
for $i = 0,1,2$, where $\g_i(\tsnr)$ is defined in (\ref{eq:gdef}).

Note that the probability of the ON state is given by
$
P_{\on}=\frac{1-p_{11}}{2-p_{11}-p_{22}}.
$
If we use the assumption $p_{11}=1-s$ and $p_{22}=s$ (and hence $P_{\on}=s$), the expression for average arrival rate can be simplified further as
\begin{gather}
r_{\avg i}^*(\tsnr, \theta_i)=\frac{s}{\theta_i}\log_e\left(\frac{e^{\theta_i C_{Ei}(\ssnr, \theta_i)} - (1-s)} {s}\right). \label{eq:ravgdiscreteq}
\end{gather}
%
\begin{figure}
\begin{center}
\includegraphics[width=0.45\textwidth]{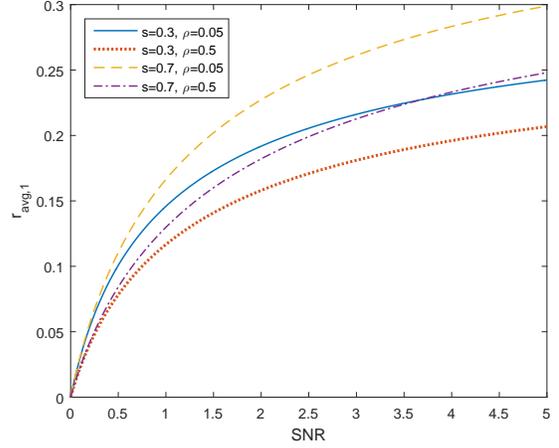}
\vspace{-0.3cm}
\caption{Maximum average arrival rate of the confidential message of the first user $r_{\avg,1}^*$ vs. average signal-to-noise ratio $\tsnr$ when $\theta_1 = 1$ and $\delta_1=0.5$.}\label{fig:ravg1SNRdisc}
\end{center}
\vspace{-0.3cm}
\end{figure}
It can be easily verified that $r_{\avg i}^*$ is a monotonic function of $s$, i.e., as $s$ (and hence ON-state probability $P_{\on}=s$) increases, the maximum average arrival rate increases. We see this effect in Fig. \ref{fig:ravg1SNRdisc} where we plot the relationship between maximum average arrival rate of the confidential message of the first user vs. average $\tsnr$ curves for different values of $s$ and correlation coefficient $\rho$. We consider a Rayleigh fading environment and assume that the fading powers $z_1$ and $z_2$ are exponentially distributed with unit means, i.e., $\E\{z_1\} = \E\{z_2\} = 1$, and correlation coefficient $\rho = \frac{\text{cov}(z_1, z_2)}{\sqrt{\text{var}(z_1)\text{var}(z_1)}}$. Numerical evaluation verifies that as $s$ increases, maximum average arrival rate increases for given $\tsnr$ and $\rho$. Hence, as the source becomes less bursty, throughput improves. Also, the correlation between the channels of the legitimate user and eavesdropper has an impact on the throughput. Higher correlation values lead to diminished secrecy capacity, which results in smaller throughput values.

\begin{figure}
\begin{center}
\includegraphics[width=0.45\textwidth]{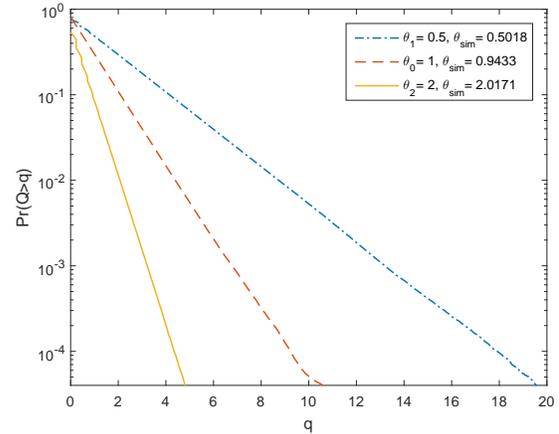}
\vspace{-0.3cm}
\caption{Buffer overflow probability $\Pr\{Q>q\}$ vs. buffer threshold $q$ for both confidential and common messages when $\theta_1=0.5$, $\theta_2=2$, $\theta_0=1$ , $\delta_1=\delta_2=0.7$, $p_{11} = p_{22} = 0.8$ and SNR = $1$. } \label{fig:bufferperfCSI}
\end{center}
\vspace{-0.3cm}
\end{figure}

We have also performed buffer simulations to further verify our theoretical analysis. Initially, we  set the values of the QoS exponent $\theta_i$, SNR, source state transition probabilities $p_{11}$ and $p_{22}$ of the ON/OFF discrete Markov source, and determined the maximum average arrival rate the system can support using the theoretical characterizations in this section. We also calculated the corresponding maximum data arrival rate $r_i$ in the ON state. Then, we initiated the simulation by generating the random data arrivals according to the Markov source model, and generating the Gaussian fading coefficients for the service rates. In this process, we have kept track of the buffer length over $10^7$ runs. We have compared the simulated buffer lengths with different thresholds to determine how frequently a threshold is exceeded and identify the overflow probabilities. In Fig. \ref{fig:bufferperfCSI}, we plot the buffer overflow probability (in logarithmic scale) vs.  buffer threshold $q$. We obtain excellent results from these simulations. Specifically, we determined the simulated QoS exponent values $\theta_{\text{sim}}$ from the slopes of the buffer overflow probability curves in the figure\footnote{Note from (\ref{eq:overflowprob-rev}) that the overflow probability is expected to behave in logarithmic scale as $\log \Pr\{Q \ge q\} \approx -\theta q + \log \varsigma$. Hence, the slope of the logarithmic overflow probability vs. buffer threshold $q$ curve is proportional to $-\theta$.}. The simulated $\theta_{\text{sim}}$ values were obtained as $2.0171, 0.9433, 0.5018$ when the corresponding theoretical $\theta$ values were $2, 1, 0.5$, respectively. Hence, if we originally set $\theta = 2$ and design the system accordingly, the buffer overflow probability decays with QoS exponent $\theta_{\text{sim}} = 2.0171$, matching the prediction very well.


\subsubsection{Markov Fluid Source} \label{subsubsec:fMarkov}
Similarly as in the case of discrete Markov source, for the ON-OFF Markov fluid source, incorporating (\ref{eq:2fluidEBW}) into \eqref{eq:equalityforQoS},
we determine the maximum average arrival rate as
\begin{align}
r_{\avg i}^*(\tsnr, \theta_i)&=P_{\on} \frac{\theta_i C_{Ei}(\ssnr, \theta_i) +\alpha+\beta} {\theta_i C_{Ei}(\ssnr, \theta_i) +\alpha} \, C_{Ei}(\tsnr, \theta_i) \nonumber
\\
&=-\frac{P_{\on}}{\theta_i} \frac{\alpha+\beta-\log_e\g_i(\tsnr)} {\alpha -\log_e\g_i(\tsnr)} \, \log_e\g_i(\tsnr) \label{eq:2fluidravg}
\end{align}
for $i = 0,1,2$.
Note that the probability of ON state is given as
$
P_{\on}=\frac{\alpha}{\alpha+\beta}.
$
\begin{figure}
\begin{center}
\includegraphics[width=0.45\textwidth]{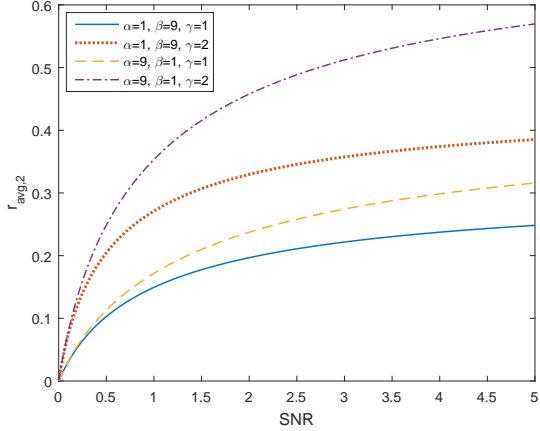}
\vspace{-0.3cm}
\caption{Maximum average arrival rate of the confidential message of the second user $r_{\avg,2}^*$ vs. average signal-to-noise ratio $\tsnr$ when $\theta_2 = 1$, $\rho=0.05$ and $\delta_2=0.5$.}\label{fig:ravg2SNRfluid}
\end{center}
\vspace{-0.3cm}
\end{figure}

In Fig. \ref{fig:ravg2SNRfluid}, we plot the maximum average arrival rate of the confidential message of the second user as a function of average $\tsnr$ while considering different channels and Markov fluid sources. Specifically, we assume different pairs of the source state transition rates $\alpha$ and $\beta$ and different expected channel gains $\E\{z_2\}=\gamma$. As in Fig. \ref{fig:ravg1SNRdisc}, we still assume that $z_1$ and $z_2$ are exponentially distributed, and $\E\{z_1\} =1$. It is observed that increasing $\alpha$ and decreasing $\beta$ simultaneously increase the ON-state probability $P_{\on}$ and reduce the burstiness of the source, and as a result, throughput increases. Furthermore, better channel conditions for the legitimate user lead to improved throughput due to increase in secrecy capacity.

\subsubsection{Discrete-Time Markov Modulated Poisson Source}
\label{subsubsec:dMMPP}
In order to express the maximum average arrival rate in terms of $C_E$, we again insert the effective bandwidth expression in (\ref{eq:discMMPPEB}) into \eqref{eq:equalityforQoS} and obtain
\begin{small}
\begin{gather}
\hspace{-.4cm}r_{\avg i}^*(\tsnr, \theta_i)\!=\!\frac{P_{\on}} {\left(e^{\theta_i}-1\right)}\!\log_e\!\left(\!\frac{1 - p_{11}\g_i(\ssnr) } {(1\!-\!p_{11}\!-\!p_{22})\g_i^2(\ssnr)\! +\!p_{22}\g_i(\ssnr) }\!\right). \label{eq:ravgdMMPP}
\end{gather}
\end{small}
\normalsize
\subsubsection{Continuous-Time Markov Modulated Poisson Source}
We find the following maximum average arrival rate $r_\avg^*$ by incorporating \eqref{eq:2MMPPEBW} into (\ref{eq:equalityforQoS}):
\begin{gather}
r_{\avg i}^*(\tsnr, \theta_i)=-\frac{P_{\on}}{\left(e^{\theta_i}-1\right)}\frac{\alpha+\beta-\log_e\g_i(\tsnr)} {\alpha -\log_e\g_i(\tsnr)} \, \log_e\g_i(\tsnr). \label{eq:2MMPPr}
\end{gather}
\section{Energy Efficiency Of Secure Transmissions with Random Data Arrivals Under QoS Constraints}

In this section, we investigate the energy efficiency of the transmission of confidential and common messages for various source types discussed previously. Using the throughput formulas we have obtained, we analyze the energy efficiency and derive closed-form expressions of the minimum energy per bit and wideband slope.

\subsection{Minimum Energy per Bit}
The minimum energy per bit in \eqref{eq:ebnomin-ra} characterizes the minimum energy needed to send one bit reliably over the wireless fading channel under statistical queueing constraints. Lower minimum energy per bit levels indicate higher energy efficiency. First, we formulate the minimum energy per bit for the confidential messages as
\begin{align}
\frac{E_b}{N_0}_{\tmin,i}=\lim_{\tsnr\rightarrow0} \frac{\delta_i \Pr(\Gamma_i) \tsnr}{r_{\avg i}^*(\tsnr,\theta_i)} = \frac{\delta_i \Pr(\Gamma_i) }{ \dot{r}_{\avg i}^*(0) }\label{eq:ebnomin_conf}
\end{align}
for $i=1,2$. Similarly for the common message, the minimum energy per bit becomes
\begin{align}
\frac{E_b}{N_0}_{\tmin,0}&=\lim_{\tsnr\rightarrow0} \frac{\left[(1-\delta_1) \Pr(\Gamma_1)+(1-\delta_2) \Pr(\Gamma_2) \right]\tsnr}{r_{\avg 0}^*(\tsnr,\theta_0)} \nonumber
 \\&= \frac{(1-\delta_1) \Pr(\Gamma_1)+(1-\delta_2) \Pr(\Gamma_2) }{ \dot{r}_{\avg 0}^*(0) }.\label{eq:ebnomin_common}
\end{align}

Below, we initially characterize the minimum energy per bit for the case of constant-rate arrivals, and subsequently show that the same minimum energy per bit levels are achieved when discrete-time Markov and Markov ON-OFF sources are considered.

\begin{prop} \label{prop:ebnocons}
When the data arrival rate is constant, the minimum energy per bit expressions for the confidential message transmissions to receivers 1 and 2 under QoS constraints are given, respectively, by
\begin{align}
{\frac{E_b}{N_0}}_{\text{min},1}= \frac{\Pr(\Gamma_1)\log_e2}{\E_{\Gamma_1}\!\!\left\{z_1- z_2\right\}} \label{eq:EbN01}
\end{align}
\begin{align}
{\frac{E_b}{N_0}}_{\text{min},2}= \frac{\Pr(\Gamma_2)\log_e2}{\E_{\Gamma_2}\left\{ z_2-z_1\right\}} \label{eq:EbN02},
\end{align}
and the minimum energy per bit for the common message transmission under QoS constraints is given by
\begin{align}
{\frac{E_b}{N_0}}_{\text{min},0}= \frac{[(1-\delta_1)\Pr(\Gamma_1)+(1-\delta_2)\Pr(\Gamma_2)]\log_e2} {(1-\delta_1)\E_{\Gamma_1}\!\!\left\{ z_2 \right\}+(1-\delta_2)\E_{\Gamma_2}\!\!\left\{z_1 \right\}} \label{eq:EbN00}
\end{align}
where $\Pr(\Gamma_1)=\Pr(z_1 < z_2)$, $\Pr(\Gamma_2)=\Pr(z_1 < z_2)$, and $\delta_i$ is fraction of the power used for the transmission of the confidential message to receiver $i$. Moreover, $\E_{\Gamma_1}$ denotes the expectation in region $\Gamma_1$ while $\E_{\Gamma_2}$ is similarly defined in the complement region $\Gamma_2$.
\end{prop}

\emph{Proof:} See Appendix \ref{subsec:ebnocons}.

When $z_1$ and $z_2$ are independent and exponentially distributed with $\E\{z_1\}=1$ and $\E\{z_2\}=\gamma$, we have $\Pr(\Gamma_1)=\frac{1}{\gamma+1}$ and $\Pr(\Gamma_2)=\frac{\gamma}{\gamma+1}$, and we can get closed-form expressions for the minimum energy per bit formulations as follows:
\begin{gather}
{\frac{E_b}{N_0}}_{\text{min},1}=\log_e2, \,\,\,\,\, {\frac{E_b}{N_0}}_{\text{min},2}=\frac{\log_e2}{\gamma} \label{eq:ebnomin-loge2}
\\
{\frac{E_b}{N_0}}_{\text{min},0}=\frac{\gamma+1}{\gamma}\log_e2.
\end{gather}


Interestingly, for both ON-OFF discrete-time Markov and Markov fluid sources,  minimum energy per bit expressions are the same as those attained in the presence of constant-rate sources.
\begin{prop} \label{prop:ebno_discrete}
When data arrivals are modeled as ON-OFF discrete-time Markov or Markov fluid processes, the minimum energy per bit expressions for confidential and common message transmissions under QoS constraints remains the same as those for the constant arrival rate model and hence are given by \eqref{eq:EbN01}, \eqref{eq:EbN02}, and \eqref{eq:EbN00}, respectively.
\end{prop}

\emph{Proof:} See Appendix  \ref{subsec:ebno_discrete}.

Heretofore, we have seen that the minimum bit energy expressions do not depend on either the queueing constraints or the source randomness. More specifically, minimum bit energy of confidential/common message transmissions are the same regardless of the value of the QoS exponent $\theta$ and whether data arrives at a constant rate or according to an ON-OFF Markov process. However, this is not the case when we consider more bursty Markov-modulated Poisson arrivals, as shown in the result below.


\begin{prop} \label{prop:ebno_MMPP}
When the source arrivals are modeled as ON-OFF discrete-time or continuous-time MMPPs, the minimum energy per bit expressions for confidential and common message transmissions under QoS constraints are given, respectively, by
\begin{align}
{\frac{E_b}{N_0}}_{\text{min},1}= \frac{(e^{\theta_1}-1)\Pr(\Gamma_1)\log_e2}{\theta_1\E_{\Gamma_1}\!\!\left\{z_1- z_2\right\}} \label{eq:EbN01MMPP}
\end{align}
\begin{align}
{\frac{E_b}{N_0}}_{\text{min},2}= \frac{(e^{\theta_2}-1)\Pr(\Gamma_2)\log_e2}{\theta_2\E_{\Gamma_2}\left\{ z_2-z_1\right\}} \label{eq:EbN02MMPP}
\end{align}
\begin{align}
{\frac{E_b}{N_0}}_{\text{min},0}= \frac{(e^{\theta_0}\!-1)[(1-\delta_1)\Pr(\Gamma_1)+(1-\delta_2)\Pr(\Gamma_2)]\log_e2} {\theta_0\left[(1-\delta_1)\E_{\Gamma_1}\!\!\left\{ z_2 \right\}+(1-\delta_2)\E_{\Gamma_2}\!\!\left\{z_1 \right\}\right]}. \label{eq:EbN00MMPP}
\end{align}

\end{prop}

\emph{Proof:} See Appendix \ref{subsec:ebno_MMPP}.

For MMPP sources, minimum energy per bit now depends on the QoS exponent through the term $\frac{e^\theta-1}{\theta}$. Since $\frac{e^\theta-1}{\theta} > 1$ for $\theta > 0$ and increases with increasing $\theta$, a higher energy per bit is required for MMPP sources (compared to constant-rate and ON-OFF Markov sources) and energy cost grows as the QoS constraints become more stringent. Interestingly, energy per bit expressions still do not depend on the specific parameters of the random arrival model (such as transition probabilities/rates of the Markov chain and intensity of the Poisson arrivals).

As also noted before, Proposition \ref{prop:ebno_discrete} shows that the minimum energy per bit for discrete-time Markov and Markov fluid sources are the same as for the constant-rate source. The primary intuitive reasoning behind this result is that the minimum energy per bit is an asymptotic performance metric achieved as $\tsnr \to 0$, and the impact of source burstiness significantly diminishes at these asymptotically low $\tsnr$ levels for discrete-time Markov and Markov fluid sources. Specifically, as $\tsnr$ diminishes, the fixed arrival rate (in the ON-state of the Markov models) that can be supported by the wireless channel decreases as well, resulting in less and less impact on buffer overflows and delay violations.

On the other hand, if the arrival process is MMPP, the intensity of the Poisson process is reduced with decreasing $\tsnr$. However, the arrival process is still a Poisson process but with a smaller intensity, meaning that there is still a probability, however small, for the instantaneous arrival rate in the ON state to be large since the arrival rate depends on the realization of a Poisson distributed random variable. Hence, MMPP source is more bursty in the low-$\tsnr$ regime than discrete-time Markov and Markov fluid sources, and this is reflected in the larger minimum energy per bit values as shown in the results of Proposition \ref{prop:ebno_MMPP}.


\subsection{Wideband Slope}
Minimum energy per bit ${\frac{E_b}{N_0}}_{\text{min}}$ is the ultimate performance limit of energy-efficient operation. At the same time, it is an asymptotic performance metric achieved in the limit as $\tsnr$ vanishes. In this subsection, we complement the ${\frac{E_b}{N_0}}_{\text{min}}\!\!\!\!-$analysis by characterizing the wideband slope of confidential and common message transmissions for different source models. Unlike the minimum energy per bit, wideband slope is distinct for each source and depends on the source statistics. In this subsection, we also provide numerical results to demonstrate the effectiveness of the linear approximation of the throughput in the low-$\tsnr$ regime in terms of ${\frac{E_b}{N_0}}_{\text{min}}$ and wideband slope $\mathcal{S}_0$, and to identify the impact of secrecy requirements, source randomness, QoS constraints, and channel correlation on energy efficiency.
\subsubsection{Constant-Rate Sources}

\begin{prop} \label{prop:widebandslope}
For constant-rate arrivals, the wideband slope expressions for common and confidential message transmissions under QoS constraint are given by
\begin{align}
\mathcal{S}_{0,i}\!=\! \frac{2\left(\E\left\{ \dot{f}_i(0)\right\}\right)^2} { \frac{\theta_i}{\log_e\!2}\text{var}\!\left(\dot{f}_i(0)\right) \!- \E\!\left\{ \ddot{f}_i(0)\right\}} \label{eq:S0i}
\end{align}
for $i=0,1,2$ where we have defined $f_i(\tsnr)=R_i(\tsnr) \log_e2$ with $R_i(\tsnr)$ being the instantaneous rate of confidential or common message given in (\ref{eq:rate0})--(\ref{eq:rate2}), and the first and second derivatives of $f_i(\tsnr)$ at $\tsnr = 0$ are given by
\begin{align}
\dot{f}_1(0)&= \delta_1 \left(z_1 - z_2\right) \ind\!\left\{z_1\!\geq \!z_2\right\}, \nonumber
\\
\dot{f}_2(0)&= \delta_2 \left( z_2-z_1 \right) \ind\!\left\{z_1\!< \!z_2\right\}, \nonumber
\\
\dot{f}_0(0)&= (1-\delta_1) z_2\ind\!\left\{z_1\!\geq \!z_2\right\}+ (1-\delta_2) z_1\ind\!\left\{z_1\!< \!z_2\right\}, \nonumber
\\
\ddot{f}_1(0)&= -\delta_1^2 \left[z_1^2 - z_2^2\right] \ind\!\left\{z_1\!\geq \!z_2\right\}, \nonumber
\\
\ddot{f}_2(0)&= -\delta_2^2 \left[ z_2^2-z_1^2 \right]\ind\!\left\{z_1\!< \!z_2\right\}, \nonumber
\\
\ddot{f}_0(0)&= -(1-\delta_1^2)z_2^2\ind\!\left\{z_1\!\geq \!z_2\right\}-(1-\delta_2^2)z_1^2 \ind\!\left\{z_1\!< \!z_2\right\}. \label{eq:f0}
\end{align}
\end{prop}

\normalsize
\emph{Proof:} See Appendix \ref{subsec:widebandslope}.

Above, $\mathcal{S}_{0,0}$ is the wideband slope for common message transmission while $\mathcal{S}_{0,1}$ and $\mathcal{S}_{0,2}$ denote the wideband slope of confidential message transmissions to receivers 1 and 2, respectively.

For independent and exponentially distributed $z_1$ and $z_2$ with $\E\{z_1\}=1$ and $\E\{z_2\}=\gamma$, the wideband slope expressions simplify to
\begin{align}
\mathcal{S}_{0,1}\!=\! \frac{2} { \frac{\theta_1}{\log_e\!2}\left(1+2\gamma\right) +4\gamma+2} \label{eq:S01}
\\
\mathcal{S}_{0,2}\!=\! \frac{2} { \frac{\theta_2}{\log_e\!2}\left(1+\frac{2}{\gamma}\right) +\frac{4}{\gamma}+2}. \label{eq:S02}
\end{align}
If we further assume that $\delta_1=\delta_2=\delta$, then the wideband slope for common message becomes
\begin{align}
\mathcal{S}_{0,0}\!=\! \frac{2} { \frac{\theta_0}{\log_e\!2}+ \frac{1-\delta^2}{(1-\delta)^2}}. \label{eq:S00}
\end{align}

\subsubsection{Discrete-Time Markov Sources}

Next, we consider ON-OFF discrete-time Markov sources with transition probabilities denoted by $p_{ij}$ for $i,j \in \{1,2\}$.

\begin{prop} \label{prop:wbsdisc}
The wideband slope expressions for confidential and common message transmissions under QoS constraint are given by
\begin{align}
\mathcal{S}_{0,i}\!=\! \frac{2\left(\E\left\{ \dot{f}_i(0)\right\}\right)^2} {\eta\frac{\theta_i}{\log_e\!2}\!\left(\E\!\left\{ \dot{f}_i(0)\right\}\right)^2\!\!\! + \! \frac{\theta_i}{\log_e\!2}\text{var}\!\left(\dot{f}_i(0)\right) \!- \E\!\left\{ \ddot{f}_i(0)\right\}} \label{eq:S0idisc}
\end{align}
for $i=0,1,2$, where $\dot{f}_i(0)$ and $\ddot{f}_i(0)$ are given in (\ref{eq:f0}). Additionally, $\eta$ above is defined as
\begin{equation}
\eta = \frac{(1-p_{22})(p_{11}+p_{22})}{(1-p_{11})(2-p_{11}-p_{22})}. \label{eq:eta}
\end{equation}
\end{prop}
\normalsize

\emph{Proof:} See Appendix \ref{subsec:wbsdisc}

Again, for independent and exponentially distributed $z_1$ and $z_2$ with $\E\{z_1\}=1$ and $\E\{z_2\}=\gamma$, the wideband slope expressions are given as
\begin{align}
\mathcal{S}_{0,1}\!=\! \frac{2} { \frac{\theta_1\eta}{\log_e\!2}+\frac{\theta_1}{\log_e\!2}\left(1+2\gamma\right) +4\gamma+2}, \label{eq:S01disc}
\\
\mathcal{S}_{0,2}\!=\! \frac{2} { \frac{\theta_2\eta}{\log_e\!2}+\frac{\theta_2}{\log_e\!2}\left(1+\frac{2}{\gamma}\right) +\frac{4}{\gamma}+2}. \label{eq:S02disc}
\end{align}
If we further assume that $\delta_1=\delta_2=\delta$, then the wideband slope for common message becomes
\begin{align}
\mathcal{S}_{0,0}\!=\! \frac{2} { \frac{\theta_0\eta}{\log_e\!2}+\frac{\theta_0}{\log_e\!2}+ \frac{1-\delta^2}{(1-\delta)^2}}. \label{eq:S00disc}
\end{align}
When compared with the corresponding wideband slope expressions in \eqref{eq:S01}--\eqref{eq:S00} for the constant-rate source, we notice that wideband slope formulas above in (\ref{eq:S01disc})--(\ref{eq:S00disc}) for the discrete Markov source differ only due to the presence of the term $\frac{\theta \eta}{\log_e2}$, which reflects essentially the source randomness with the parameter $\eta$. This additional term leads to smaller wideband slopes, indicating the detrimental impact of source randomness on energy efficiency. Note also that when $p_{11} = 0$ and $p_{22} = 1$, discrete Markov essentially becomes a constant-rate source and we have $\eta = 0$.
\begin{figure}
\begin{center}
\includegraphics[width=0.45\textwidth]{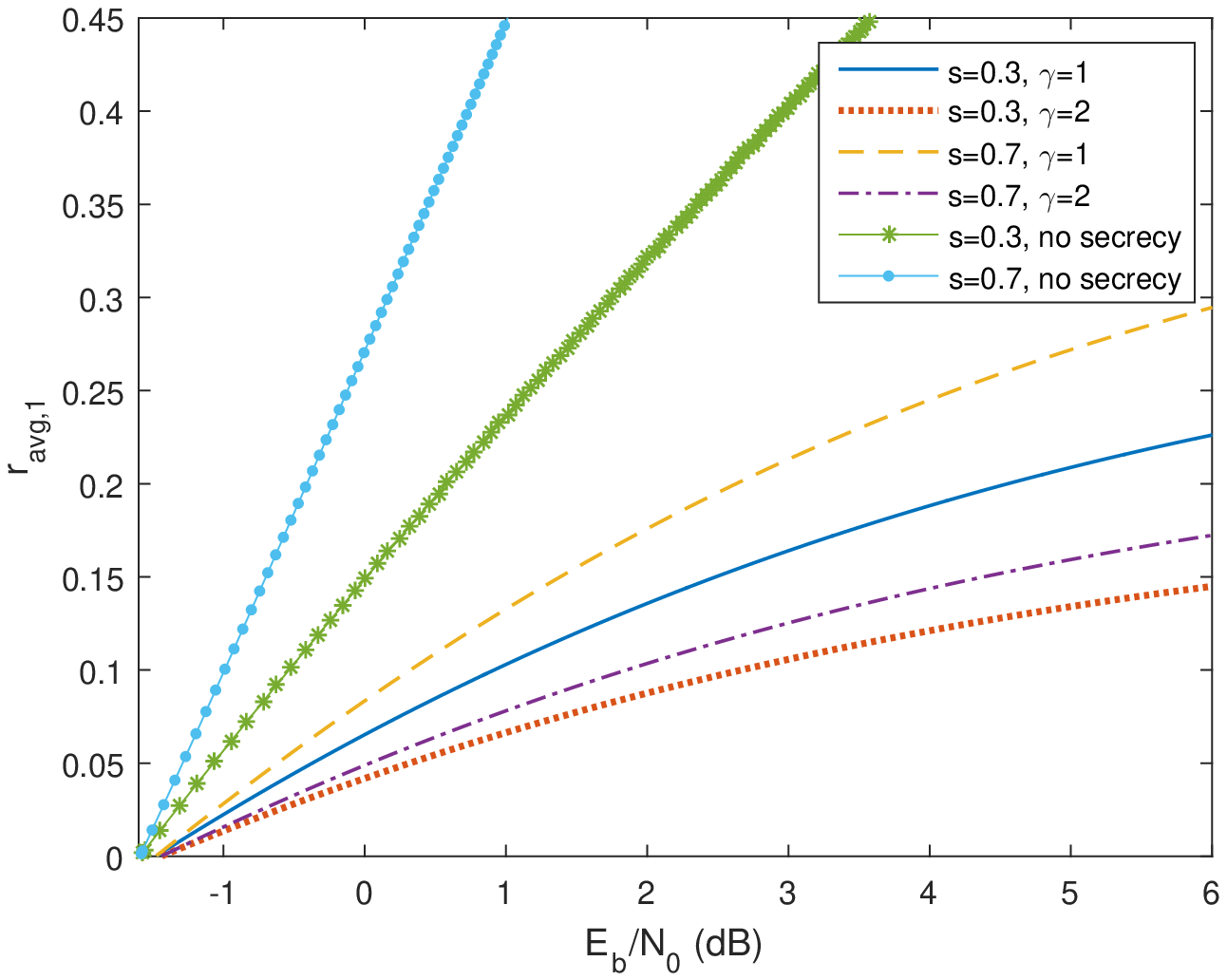}
\vspace{-0.3cm}
\caption{Maximum average arrival rate of first user's confidential message $r_{\avg1}^*$ vs. energy per bit $\frac{E_b}{N_0}$ in dB when $\theta_1 = 1$, $\rho=0.05$ and $\delta_1=0.5$.}\label{fig:ravg1EbN0disc}
\end{center}
\vspace{-0.3cm}
\end{figure}

In Fig. \ref{fig:ravg1EbN0disc}, the maximum average arrival rate of the confidential message for the first user vs. energy per bit is plotted. We consider an ON-OFF discrete Markov source with $p_{11}=1-s$ and $p_{22}=s$ (and hence $P_{\on}=s$). We assume $\theta=1$ and $\delta_1=0.5$. The channel power gains $z_1$ and $z_2$ are exponentially distributed with  $\E\{z_1\}=1$, $\E\{z_2\}=\gamma$ and correlation coefficient $\rho=0.05$. As predicted, the minimum energy per bit does not depend on source burstiness or the second user channel statistics, i.e., $\gamma$. There is a slight increase in the minimum energy per bit values achieved in the cases of secrecy as compared to no secrecy. The main reason for this is the correlation in the channel conditions of the two users. Without any correlation, the minimum energy per bit becomes equal to $-1.59$ dB. As a result of similar minimum energy per bit values, wideband slope becomes a critical performance indicator in the low-SNR regime. We notice that wideband slope diminishes when secrecy requirements are imposed and also when source burstiness increases with diminishing  ON-state probability $P_{\on}=s$. We also observe that, as the second user (or equivalently eavesdropper) channel conditions improve, i.e., as $\gamma$ increases, we have smaller wideband slopes.

\begin{figure}
\begin{center}
\includegraphics[width=0.45\textwidth]{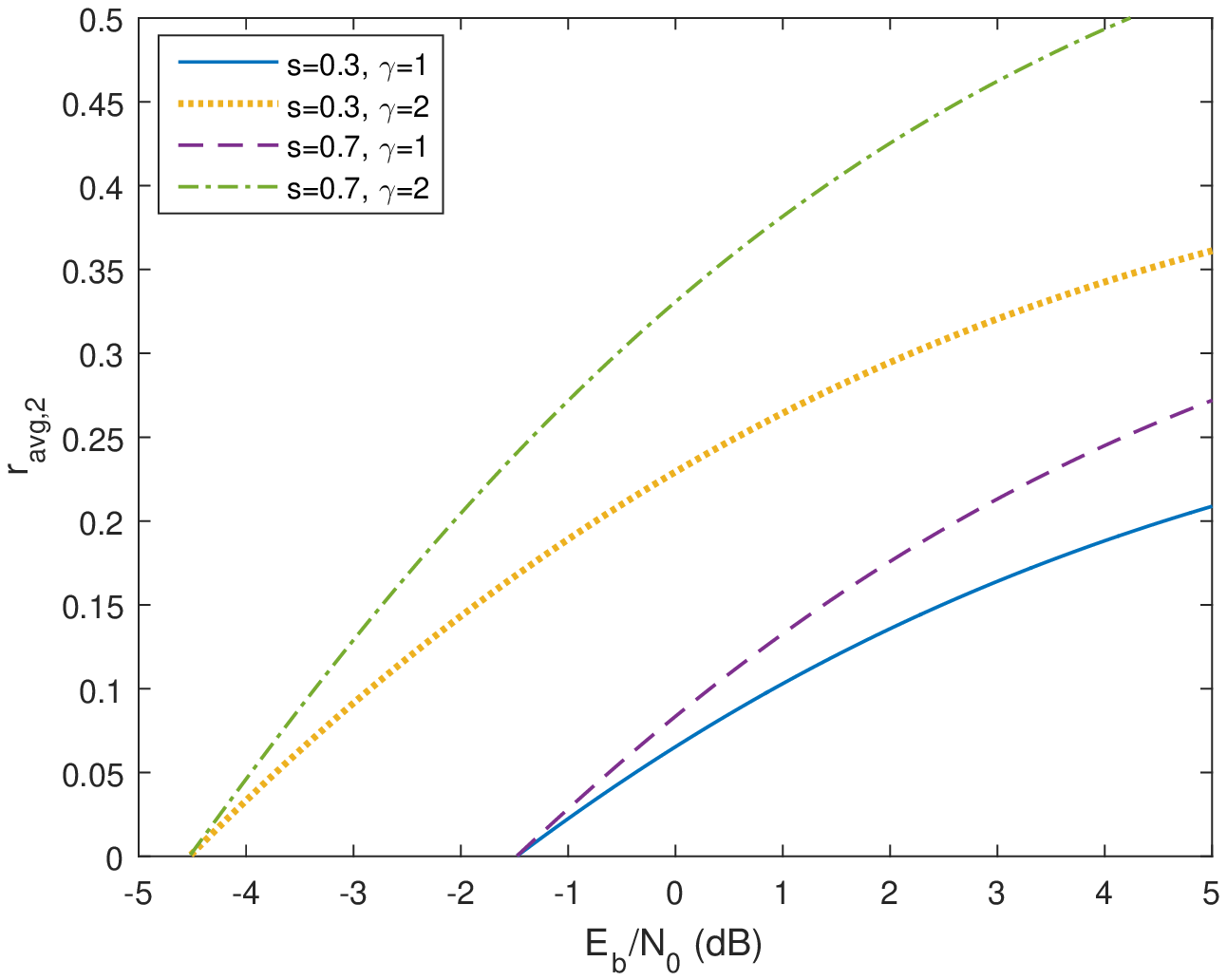}
\vspace{-0.3cm}
\caption{Maximum average arrival rate of second user's confidential message $r_{\avg2}^*$ vs. energy per bit $\frac{E_b}{N_0}$ in dB when $\theta_2 = 1$, $\rho=0.05$ and $\delta_2=0.5$.}\label{fig:ravg2EbN0disc}
\end{center}
\vspace{-0.3cm}
\end{figure}

In Fig. \ref{fig:ravg2EbN0disc}, the maximum average arrival rate of the confidential messages for the second user vs. energy per bit is plotted. Similarly as before, we set $\theta=1$, $\rho=0.05$ and $\delta_2=0.5$. Again, the minimum energy per bit does not depend on source burstiness. On the other hand, we observe that wideband slope increases as source becomes less bursty, i.e., as $q$ increases. Also, better channel conditions for the legitimate user (i.e., larger $\gamma$) increases the energy efficiency.

\begin{figure}
\begin{center}
\includegraphics[width=0.45\textwidth]{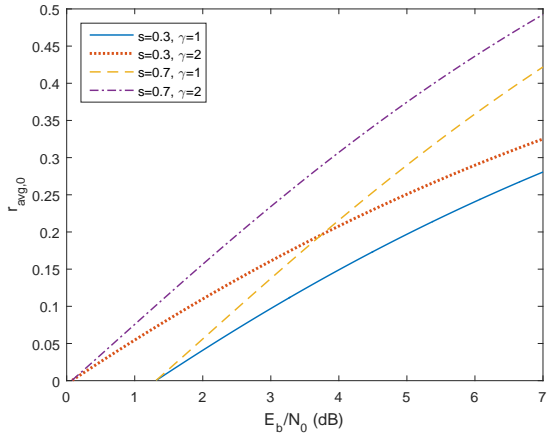}
\vspace{-0.3cm}
\caption{Maximum average arrival rate of common message $r_{\avg0}^*$ vs. energy per bit $\frac{E_b}{N_0}$ in dB when $\theta_0 = 1$, $\rho=0.05$ and $\delta_1=\delta_2=0.5$.}\label{fig:ravg0EbN0disc}
\end{center}
\vspace{-0.3cm}
\end{figure}

We illustrate the spectral efficiency curve for the common message in Fig. \ref{fig:ravg0EbN0disc}, assuming the parameter setting $\theta = 1$, $\rho=0.05$ and $\delta_1=\delta_2=0.5$. We again verify that source characteristics do not play a role in the value of the minimum energy per bit. Better channel conditions for the second user improve the overall energy efficiency of the transmission of the common message by improving the minimum energy per bit. We also notice that wideband slope is the same when we alter the channel conditions. However, source burstiness has a negative impact on the wideband slope, thus, on the energy efficiency as well.

\subsubsection{ Markov Fluid Sources}

In the following, we characterize the wideband slope in the case of ON-OFF Markov fluid arrivals with transition rates $\alpha$ and $\beta$.

\begin{prop} \label{prop:wbsfluid}
The wideband slope expressions for confidential and common message transmissions under QoS constraint are given by
\begin{align}
\mathcal{S}_{0,i}\!=\! \frac{2\left(\E\left\{ \dot{f}_i(0)\right\}\right)^2} {\zeta\frac{\theta_i}{\log_e\!2}\!\left(\E\!\left\{ \dot{f}_i(0)\right\}\right)^2\!\!\! + \! \frac{\theta_i}{\log_e\!2}\text{var}\!\left(\dot{f}_i(0)\right) \!+ \E\!\left\{ \ddot{f}_i(0)\right\}} \label{eq:S0ifluid}
\end{align}
for $i=0,1,2$, where $\dot{f}_i(0)$ and $\ddot{f}_i(0)$ are defined in \eqref{eq:f0}. Note further that $\zeta$ is defined as
\begin{equation}
\zeta=\frac{2\beta}{\alpha(\alpha+\beta)}. \label{eq:zeta}
\end{equation}
\end{prop}
\normalsize
\emph{Proof: } See Appendix \ref{subsec:wbsfluid}.

Similarly as for the previous arrival models, we can simplify the wideband expressions for the confidential message transmissions to the following when we have independent and exponentially distributed $z_1$ and $z_2$ with $\E\{z_1\}=1$ and $\E\{z_2\}=\gamma$:
\begin{align}
\mathcal{S}_{0,1}\!=\! \frac{2} { \frac{\theta_1\zeta}{\log_e\!2}+\frac{\theta_1}{\log_e\!2}\left(1+2\gamma\right) +4\gamma+2}, \label{eq:S01fluid}
\\
\mathcal{S}_{0,2}\!=\! \frac{2} { \frac{\theta_2\zeta}{\log_e\!2}+\frac{\theta_2}{\log_e\!2}\left(1+\frac{2}{\gamma}\right) +\frac{4}{\gamma}+2}. \label{eq:S02fluid}
\end{align}
If we further assume that $\delta_1=\delta_2=\delta$, then the wideband slope for common message becomes
\begin{align}
\mathcal{S}_{0,0}\!=\! \frac{2} { \frac{\theta_0\zeta}{\log_e\!2}+\frac{\theta_0}{\log_e\!2}+ \frac{1-\delta^2}{(1-\delta)^2}}. \label{eq:S00fluid}
\end{align}
The common theme in the above expressions and the ones corresponding to other source types (i.e., expressions in (\ref{eq:S01})--(\ref{eq:S00}) and (\ref{eq:S01disc})--(\ref{eq:S00disc})) is that wideband slope expressions depend on three critical factors: QoS exponent $\theta$, source burstiness parameter ($\zeta$ in the case of Markov fluid source and $\eta$ in the case discrete Markov source, which both become zero when the arrival rate is constant), and channel statistics through $\E\{z_2\} = \gamma$. For instance, wideband slopes diminish as $\theta$ increases and more stringent buffer/delay constraints are imposed.

\begin{figure}
\begin{center}
\includegraphics[width=0.45\textwidth]{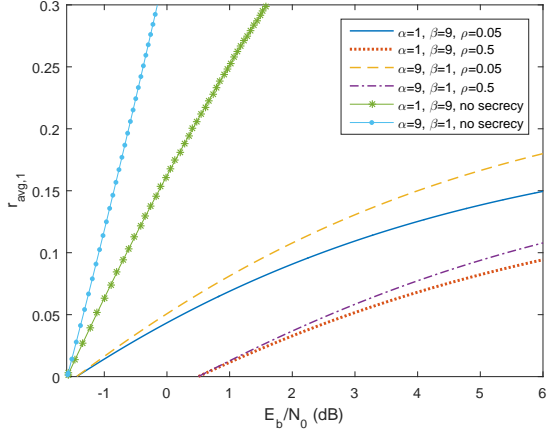}
\vspace{-0.3cm}
\caption{Maximum average arrival rate of first user's confidential message $r_{\avg1}^*$ vs. energy per bit $\frac{E_b}{N_0}$ when $\theta_1 = 1$, $\gamma=1$ and $\delta_1=0.5$.}\label{fig:ravg1EbN0fluid}
\end{center}
\vspace{-0.3cm}
\end{figure}

We depict, in Fig. \ref{fig:ravg1EbN0fluid}, the maximum average arrival rate of the confidential message for the first user vs. energy per bit for Markov fluid sources with different values of $\alpha$ and $\beta$. We assume $\theta = 1$, $\gamma=1$ and $\delta_1=0.5$. In the case of no secrecy, the minimum energy per bit is equal to $-1.59$ dB and it remains unchanged under different source characteristics. With secrecy, source burstiness again does not impact the minimum energy per bit. However, as channel correlation increases, the energy efficiency degrades due to higher minimum energy per bit. Additionally, the source characteristics have significant impact on the wideband slope e.g. wideband slope decreases as source becomes more bursty (i.e., as we change the state transition rates from $\alpha=9$ and $\beta=1$ to $\alpha=1$ and $\beta=9$).

\begin{figure}
\begin{center}
\includegraphics[width=0.45\textwidth]{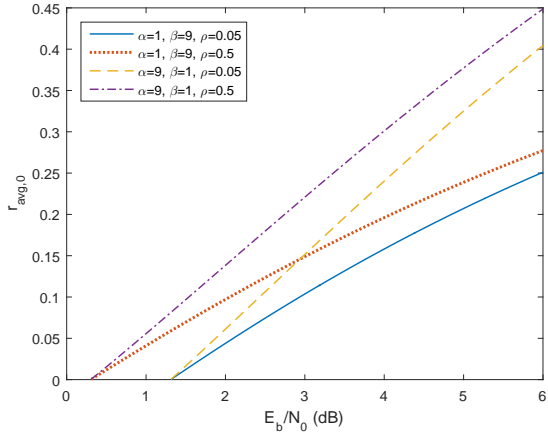}
\vspace{-0.3cm}
\caption{Maximum average arrival rate of common message $r_{\avg0}^*$ vs. energy per bit $\frac{E_b}{N_0}$ when $\theta_0 = 1$, $\gamma=1$ and $\delta_1=\delta_2=0.5$.}\label{fig:ravg0EbN0fluid}
\end{center}
\vspace{-0.3cm}
\end{figure}

In Fig. \ref{fig:ravg0EbN0fluid}, the maximum average arrival rate of the common messages for the first user vs. energy per bit for the source with Markov fluid characteristics is illustrated. We assume $\theta = 1$, $\gamma=1$ and $\delta_1=\delta_2=0.5$. The minimum energy per bit only changes for the different values of correlation. Interestingly, in this case the energy efficiency gets better with increasing correlation. The intuition behind this is that the common message throughput is limited by the worst of the channels of the first and second users. As correlation increases, the discrepancy between the conditions of the channels is reduced, improving the throughput. Additionally, wideband slope is higher for less bursty systems.

\subsubsection{Discrete-Time MMPP Sources}

Next, we address ON-OFF discrete-time MMPP sources.

\begin{prop} \label{prop:wbsMMPPdisc}
The wideband slope expressions for confidential and common message transmissions under QoS constraint are given by
\begin{align}
\mathcal{S}_{0,i}\!=\! \frac{\frac{2\theta_i}{e^{\theta_i}-1}\left(\E\left\{ \dot{f}_i(0)\right\}\right)^2} {\eta\frac{\theta_i}{\log_e\!2}\!\left(\E\!\left\{ \dot{f}_i(0)\right\}\right)^2\!\!\! + \! \frac{\theta_i}{\log_e\!2}\text{var}\!\left(\dot{f}_i(0)\right) \!+ \E\!\left\{ \ddot{f}_i(0)\right\}}. \label{eq:S0i}
\end{align}
for $i=0,1,2$ where $\dot{f}_i(0)$ and $\ddot{f}_i(0)$ are defined in \eqref{eq:f0} and $\eta$ is defined in \eqref{eq:eta}.
\end{prop}
\normalsize
We omit the proof as it is rather straightforward due to the relationship between the throughputs of the discrete Markov source and the discrete MMPP source.
\begin{figure}
\begin{center}
\includegraphics[width=0.45\textwidth]{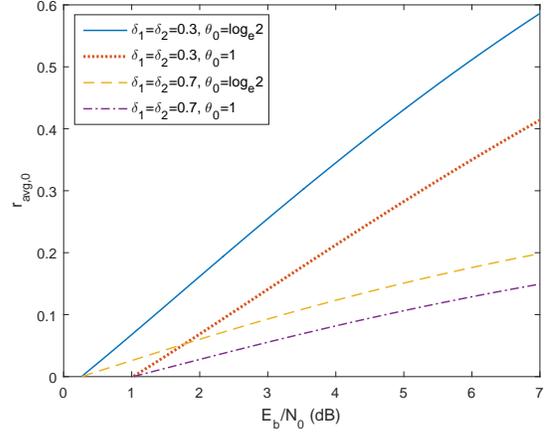}
\vspace{-0.3cm}
\caption{Maximum average arrival rate of common message $r_{\avg0}^*$ vs. energy per bit $\frac{E_b}{N_0}$ in dB when $\rho=0.8$, $\gamma=1$, $p_{11}=0.1$ and $p_{22}=0.9$.}\label{fig:ravg0EbN0MMPPdisc}
\end{center}
\vspace{-0.3cm}
\end{figure}

In Fig. \ref{fig:ravg0EbN0MMPPdisc}, we illustrate the maximum average arrival rate of the common message vs. energy per bit when the source is ON-OFF discrete-time MMPP. We set $\rho=0.8$, $\gamma=1$, $p_{11}=0.1$ and $p_{22}=0.9$, and study the impact of different values of $\theta_0$ and $\delta_i$. For the MMPP source, the minimum energy per bit depends on the QoS exponent $\theta_0$ and it will improve when $\theta_0$ decreases, indicating less stringent queueing constraints. Power allocation has no impact on the minimum energy per bit. However, with more power allocated to the common message, the wideband slope becomes higher.

\subsubsection{Continuous-Time MMPP Sources}

Finally, we consider continuous-time MMPP sources.

\begin{prop} \label{prop:wbsMMPPfluid}
The wideband slope expressions for confidential and common message transmissions under QoS constraint are given by
\begin{align}
\mathcal{S}_{0,i}\!=\! \frac{\frac{2\theta_i}{e^{\theta_i}-1}\left(\E\left\{ \dot{f}_i(0)\right\}\right)^2} {\zeta\frac{\theta_i}{\log_e\!2}\!\left(\E\!\left\{ \dot{f}_i(0)\right\}\right)^2\!\!\! + \! \frac{\theta_i}{\log_e\!2}\text{var}\!\left(\dot{f}_i(0)\right) \!+ \E\!\left\{ \ddot{f}_i(0)\right\}} \label{eq:S0i}
\end{align}
for $i=0,1,2$ where $\dot{f}_i(0)$ and $\ddot{f}_i(0)$ are defined in \eqref{eq:f0} and $\zeta$ is defined in \eqref{eq:zeta}.
\end{prop}
\normalsize
We again omit the proof due to the fact that the result readily follows from the relationship between the throughputs of Markov fluid and fluid MMPP sources.
\begin{figure}
\begin{center}
\includegraphics[width=0.45\textwidth]{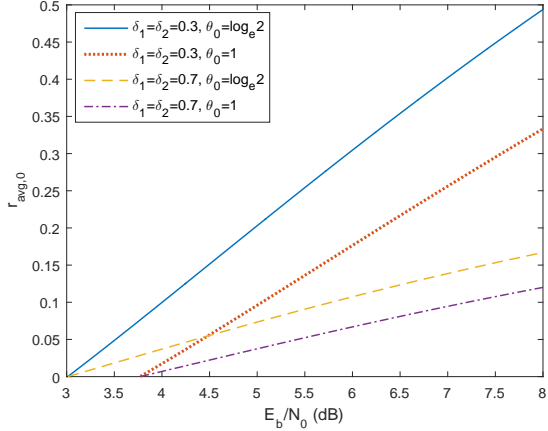}
\vspace{-0.3cm}
\caption{Maximum average arrival rate of common message $r_{\avg0}^*$ vs. energy per bit $\frac{E_b}{N_0}$ in dB when $\rho=0$, $\gamma=1$, $\alpha=9$ and $\beta=1$.}\label{fig:ravg0EbN0MMPPfluid}
\end{center}
\vspace{-0.3cm}
\end{figure}

In Fig. \ref{fig:ravg0EbN0MMPPfluid}, we illustrate the maximum average arrival rate of the common message vs. energy per bit when the source is ON-OFF continuous-time MMPP. We assume $\rho=0$, $\gamma=1$, $\alpha=9$ and $\beta=1$. We again study the impact of $\theta_0$ and $\delta_i$ on energy efficiency similarly as in Fig. \ref{fig:ravg0EbN0MMPPdisc} but the main difference is that there is no correlation in Fig. \ref{fig:ravg0EbN0MMPPfluid}. As in the previous discussion, the minimum energy per bit improves with lower $\theta_0$ values, and increasing the power allocation on common message transmission increases the wideband slope, and thus improves the energy efficiency. Interestingly, when compared with Fig. \ref{fig:ravg0EbN0MMPPdisc}, we notice that having no correlation between the channels of the two users hurts the energy efficiency of the common message transmission as it increases the minimum energy per bit significantly.

\section{Throughput and Energy Efficiency with no Channel Knowledge at the Transmitter} \label{section:noCSI}

In this section, we depart from the perfect transmitter CSI assumption of the previous sections and consider a scenario in which the transmitter has no CSI. Specifically, we assume that the transmitter does not know the realizations of the channel fading coefficients, which is relevant in cases in which the eavesdropper is passive and malicious. This also represents a worst-case scenario due to the fact that even the legitimate channel is not known. Treating the eavesdropper as malicious, we address a special case of the previously treated system model. In particular, we do not consider common message transmission and assume that the transmitter just intends to send confidential messages to receiver 1 while keeping them private from receiver 2 (which is regarded as the eavesdropper).

Not knowing the realizations of the channel fading coefficients $h_1$ and $h_2$, the transmitter sends the data at the fixed rate of $\lambda$ bits/s/Hz. As before, instantaneous secrecy capacity $R(\tsnr)  =\left[\log_2(1 + \tsnr z_1)-\log_2(1 + \tsnr z_2)\right]^+$  quantifies the maximum achievable rates of secure communication where $z_i  = |h_i|^2$. Hence, if $\lambda \le R(\tsnr)$, then reliable and secure communication is attained and therefore the transmitted message is decoded correctly while eavesdropper is being kept ignorant of the message. If, on the other hand, $\lambda > R(\tsnr)$, secrecy outage occurs. Under these assumptions, the wireless link can be modeled as a two-state discrete-time Markov chain. Specifically, the channel is assumed to be in the ON state if $\lambda \le R(\tsnr)$, while the channel is in the OFF state when $\lambda > R(\tsnr)$. The steady-state probability for the ON state can be easily obtained as
\begin{align}
\p\{\Gamma\}&=\p\{R(\tsnr)>\lambda\}=\p\left\{z_1 > 2^\lambda z_2+ \frac{2^\lambda-1}{\tsnr}\right\}
\\
&=\int_0^\infty \int_{2^\lambda z_2+ \frac{2^\lambda-1}{\ssnr}}^\infty p(z_1, z_2)dz_1 dz_2 \label{eq:probgamma}
\end{align}
where we define $\Gamma=\left\{(z_1, z_2) \in \mathbb{R}^+ : \lambda < R(\tsnr)\right\}$.

\subsection{Effective Capacity with no Channel Knowledge at the Transmitter}

In \cite[Chap. 7, Example 7.2.7]{Changbook}, it is shown for Markov modulated processes that
\begin{gather}
\label{eq:theta-envelope}
\frac{\Lambda(\theta)}{\theta} = \frac{1}{\theta} \log_e \E\{ \rho\big(\phi(\theta)\M\big)\}.
\end{gather}
Above, $\M$ is the transition matrix of the underlying Markov process, and $\phi(\theta)$ is a diagonal matrix whose components are the moment generating functions of the processes in the Markov states.
We assume that the fading coefficients $\{h_i\}$ change independently from one block to another. Under this assumption, the effective capacity can be obtained as
\begin{align} \label{eq:impcsiCE}
C_E(\tsnr,\theta) = -\frac{\Lambda(-\theta)}{\theta} = -\frac{1}{\theta}\log_e\left[1-\p\{\Gamma_1\}\left(1-e^{-\theta \lambda}\right)\right]
\end{align}
where $\p\{\Gamma\}$ is the channel ON-state probability given in \eqref{eq:probgamma}.

%
%

\subsection{Energy Efficiency with Discrete Markov Sources}

First, we consider ON-OFF discrete Markov sources. We also assume that channel fading powers $z_1$ and $z_2$ are independent exponentially distributed with means 1 and $\gamma$, respectively.
In the following result, we characterize the considered energy efficiency metrics under these assumptions.

\begin{prop}\label{prop:impcsi}
The minimum energy per bit and wideband slope achieved with fixed-rate secure transmissions in the presence of an eavesdropper with ON-OFF discrete Markov data arrivals and statistical QoS constraints are given by
\begin{align}
\frac{E_b}{N_0}_{\tmin}=e(\gamma+1)\log_e2, \quad \text{ and} \label{eq:impcsiebno}
\\
\mathcal{S}_{0}\!=\frac{1}{\frac{\theta (\eta-1)}{2\log_e2}+\frac{\theta e (\gamma+1) }{2\log_e2}+ e\gamma+\frac{e (\gamma+1)}{2} }, \label{eq:impcsiS0}
\end{align}
respectively, with $\eta$ defined in \eqref{eq:eta}.
\end{prop}
\normalsize

\emph{Proof: } See Appendix \ref{subsec:impcsi}.

As in the perfect CSI case, the minimum energy per bit in \eqref{eq:impcsiebno} does not depend on the QoS exponent $\theta$ and source statistics while the wideband slope in \eqref{eq:impcsiS0} depends on both. Specifically, wideband slope decreases with stricter QoS limitations (i.e., with increasing $\theta$) and increased source burstiness (i.e., with larger $\eta$).

It is also interesting to compare the minimum energy per bit expressions achieved with perfect CSI and no CSI. Recall from (\ref{eq:ebnomin-loge2}) that with perfect CSI, the minimum energy per bit for the confidential message transmission to receiver 1 assuming exponentially distributed fading powers with $\E\{z_1\} = 1$ and $\E\{z_2\} = \gamma$ is
\begin{gather}
\frac{E_b}{N_0}_{\tmin} = \log_e2.
\end{gather}
Comparing this with (\ref{eq:impcsiebno}), we immediately identify the additional energy cost per bit of not having channel knowledge at the transmitter as $\left[e(\gamma+1)-1\right] \log_e2$. Hence, the characterization in Proposition \ref{prop:impcsi} nicely quantifies the energy cost of not having transmitter CSI in secure wireless transmissions.

\begin{figure}
\begin{center}
\includegraphics[width=0.45\textwidth]{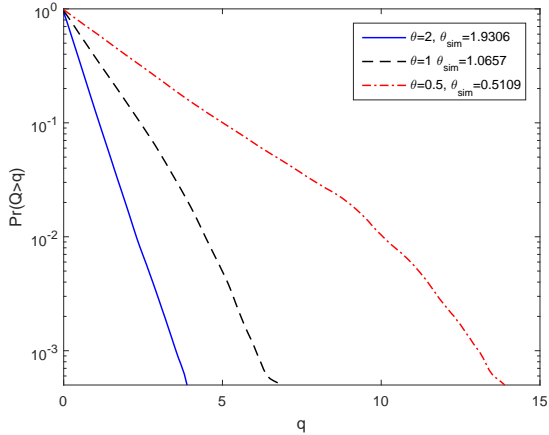}
\vspace{-0.3cm}
\caption{Buffer overflow probability $\Pr\{Q>q\}$ vs. buffer threshold $q$ for different values of $\theta$. $p_{11} = p_{22} = 0.8$, SNR = $0.05$ } \label{fig:bufferqueue}
\end{center}
\vspace{-0.3cm}
\end{figure}

Following the same methodology as described in the discussion of Fig. \ref{fig:bufferperfCSI}, we have again performed simulations in the case of no transmitter CSI. In Fig. \ref{fig:bufferqueue}, we plot the buffer overflow probability vs.  buffer threshold $q$. We again have very good agreement with theoretical predictions. In particular, the simulated $\theta_{\text{sim}}$ values were obtained as $1.9306, 1.0657, 0.5109$ when the corresponding theoretical $\theta$ values were $2, 1, 0.5$, respectively.


\begin{figure}
\begin{center}
\includegraphics[width=0.45\textwidth]{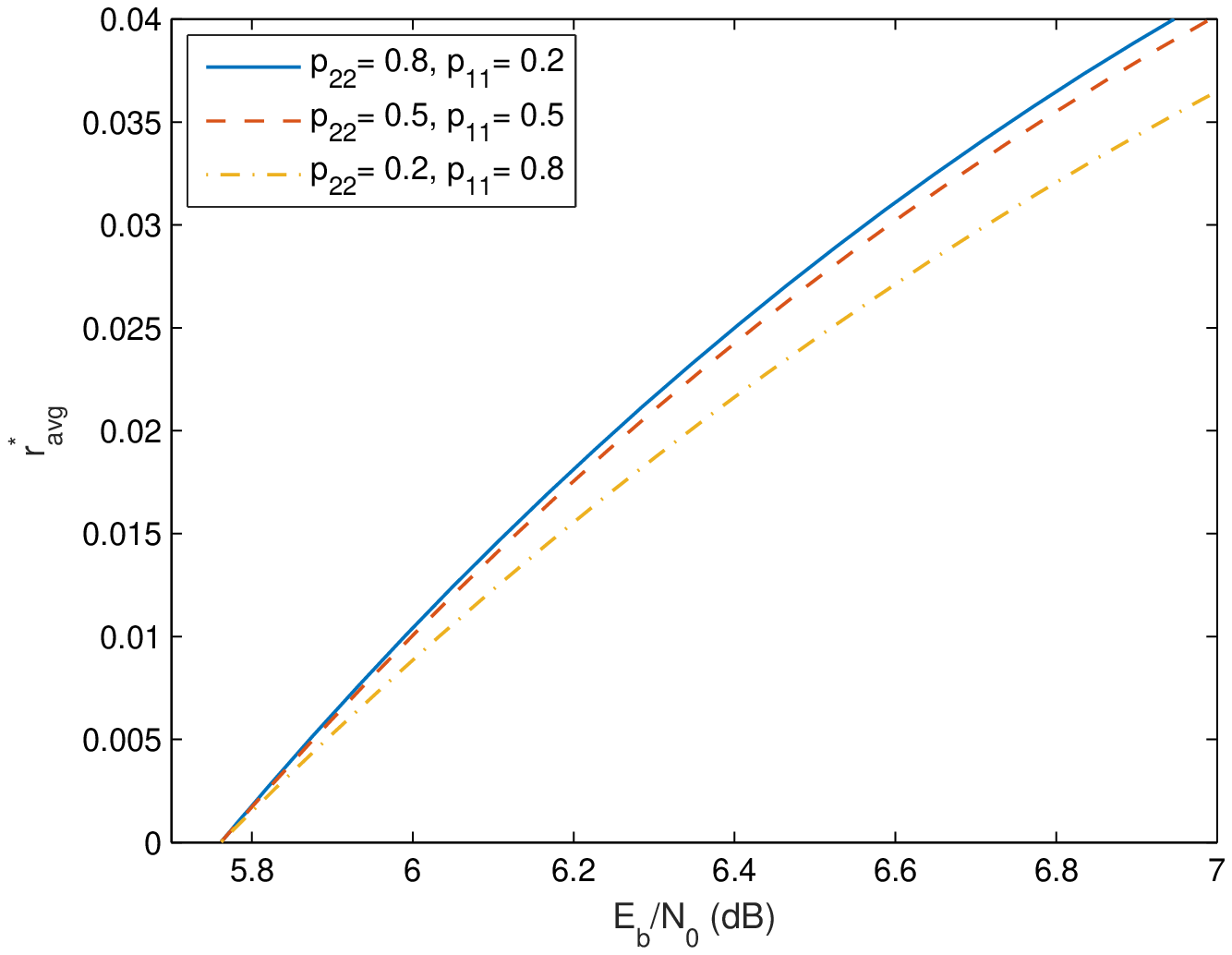}
\vspace{-0.3cm}
\caption{Maximum average arrival rate $ r_{\avg}^*$ vs. energy per bit $\frac{E_b}{N_0}$ with various source statistics when $\theta = 0.5$.}\label{fig:impcsidisc}
\end{center}
\vspace{-0.3cm}
\end{figure}

As also noted above, Proposition \ref{prop:impcsi} shows that while the minimum energy per bit does not depend on the source statistics and QoS exponent $\theta$, the wideband slope depends on both and decreases as burstiness parameter $\eta$ increases. We see these clearly in Fig. \ref{fig:impcsidisc}, where we plot the maximum average arrival rate vs. energy per bit for discrete Markov sources with varying statistics. As predicted, the minimum energy per bit stays same at $5.76$ dB, which is more than 7 dB larger than the minimum energy per bit of $-1.59$ dB achieved in the case of perfect CSI. We also observe that source with smaller $p_{11}$ and greater $p_{22}$ (while keeping $p_{11} + p_{22} = 1$) has a smaller $\eta$ value and  correspondingly larger wideband slope. Hence, lower source burstiness benefits the energy efficiency.


\subsection{Energy Efficiency with Markov Fluid Sources}

In this section, we consider ON-OFF Markov fluid sources and similarly as in the previous section identify the energy efficiency metrics.

\begin{prop}\label{prop:impcsifluid}
The minimum energy per bit and wideband slope achieved with fixed-rate secure transmissions in the presence of an eavesdropper with ON-OFF Markov fluid data arrivals and statistical QoS constraints are given by
\begin{align}
\frac{E_b}{N_0}_{\tmin}=e(\gamma+1)\log_e2, \quad \text{ and} \label{eq:impcsiebnofluid}
\\
\mathcal{S}_{0}\!=\frac{1}{\frac{\theta (\zeta-1)}{2\log_e2}+\frac{\theta e (\gamma+1) }{2\log_e2}+ e\gamma+\frac{e (\gamma+1)}{2} }, \label{eq:impcsiS0fluid}
\end{align}
respectively, where $\zeta$ is defined in \eqref{eq:zeta}

\end{prop}

\emph{Proof:} See Appendix \ref{subsec:impcsifluid}.

\begin{figure}
\begin{center}
\includegraphics[width=0.45\textwidth]{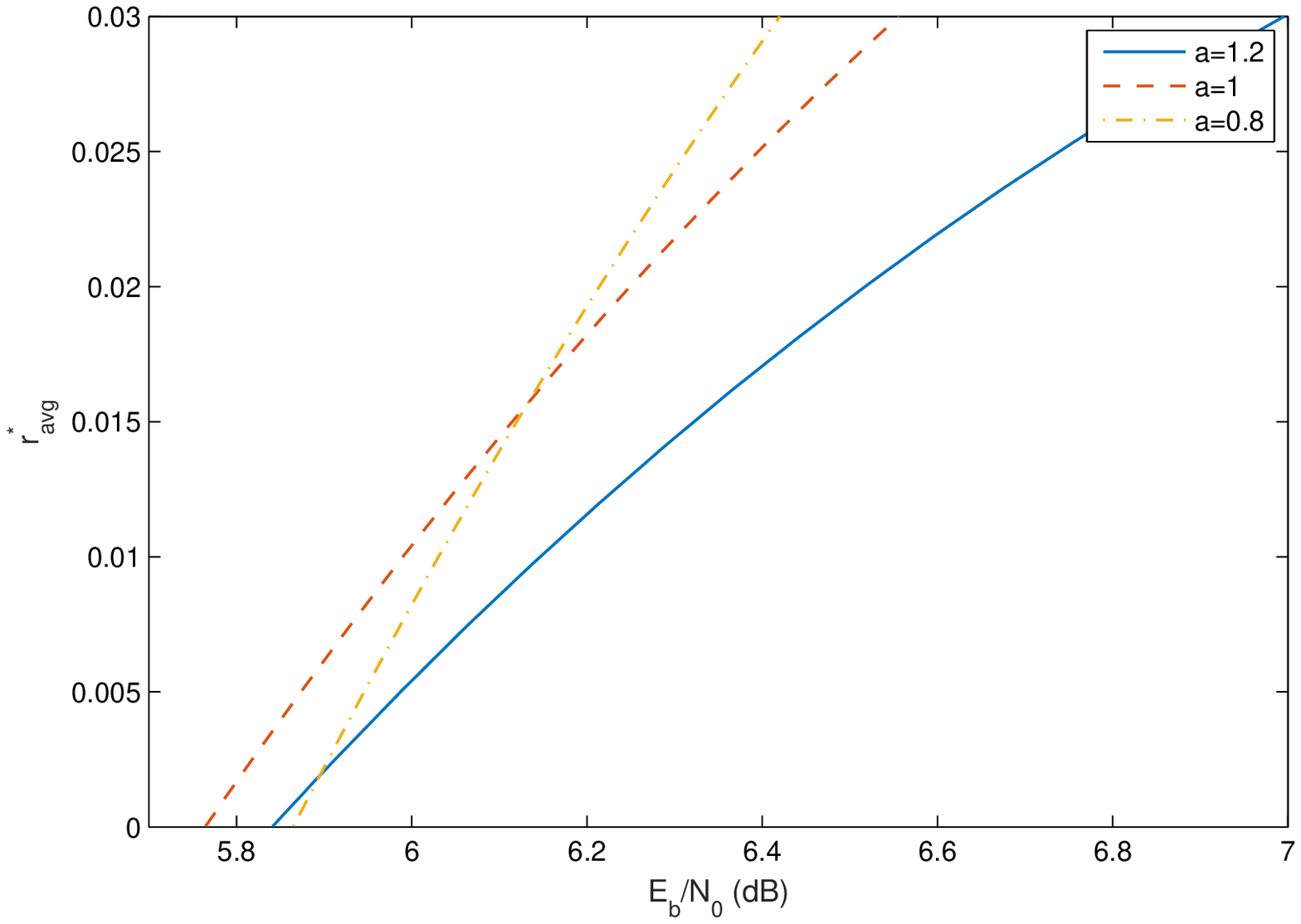}
\vspace{-0.3cm}
\caption{Maximum average arrival rate $ r_{\avg}^*$ vs. energy per bit $\frac{E_b}{N_0}$ with various source statistics when $\theta = 0.5$.}\label{fig:impcsifluid}
\end{center}
\vspace{-0.3cm}
\end{figure}

In Fig. \ref{fig:impcsifluid}, we depict the maximum average arrival rate vs. energy per bit curves for Markov fluid sources. We change the fixed the rate parameter $a$ (introduced in (\ref{eq:lambdataylor})) and compare the curves. As expected, the optimal selection of $a=1$ returns the smallest value for the minimum energy per bit. However, as $\tsnr$ increases, higher performance is achieved when operating with $a=0.8$, indicating that $a = 1$ is optimal only at sufficiently small values of $\tsnr$.


\section{Conclusion}\label{sec:Conc}
In this paper, we have analyzed the throughput and energy efficiency of secure broadcast transmissions of two confidential messages and multicast transmission of a common message to two users under statistical delay/buffer QoS constraints. Considering Markovian data arrivals to the buffers, we have identified the maximum average arrival rates (or equivalently the throughput) and have determined the minimum energy per secret bit for confidential message transmission and minimum energy per bit for common message transmission. We have also obtained a linear approximation of the maximum average arrival rates in terms of $\frac{E_b}{N_0}$ by identifying the wideband slope.

The key observations and results are the following. Secure throughput is shown to decrease as the source becomes more bursty or the channel correlation increases. Throughput also diminishes when more stringent QoS requirements (indicated by higher values of the QoS exponent $\theta$) are imposed. In terms of energy efficiency, correlation works in favor of common message transmission while it works against the confidential message transmission. We have seen that in general, security requirements, source burstiness, and QoS constraints increase energy requirements. This is due to the facts that security considerations increase minimum energy per bit, and QoS constraints and source burstiness, while not having an impact on the minimum energy per bit in the cases of discrete Markov and Markov fluid sources, reduce the wideband slope. We have also noted that the more bursty MMPP sources require minimum energy per bit values that depends on the QoS exponent $\theta$ and increases with increasing $\theta$. Hence, in this case, energy requirements grow significantly as buffer/delay constraints become stricter.

Finally, assuming no instantaneous channel knowledge at the transmitter and fixed-rate transmissions, we have identified the throughput and energy efficiency expressions for both discrete discrete Markov and Markov fluid sources. Via these characterizations, we have identified the additional energy costs due to not knowing the channel.

\appendix

\subsection{Proof of Proposition \ref{prop:ebnocons}} \label{subsec:ebnocons}


First, we define minimum energy per bit for the confidential messages as
\begin{align}
\frac{E_b}{N_0}_{\tmin,i}= \frac{\delta_i \Pr(\Gamma_i) }{ \dot{r}_{\avg i}^*(0) }\label{eq:ebnomin_conf2}
\end{align}
where $i=1,2$. Similarly for the common message, the minimum energy per bit becomes
\begin{align}
\frac{E_b}{N_0}_{\tmin,0}= \frac{(1-\delta_1) \Pr(\Gamma_1)+(1-\delta_2) \Pr(\Gamma_2) }{ \dot{r}_{\avg 0}^*(0) }.\label{eq:ebnomin_common2}
\end{align}
As the arrival rate is constant, we can use effective capacity as the throughput formula. Therefore, we can exchange $\dot{r}_\avg^*(0)$ with $\dot C_{E}(0)$ in the minimum energy per bit equation. For the proofs, we primarily focus on the $\g(\tsnr)$ function that is defined in \eqref{eq:gdef}.

Now, the first derivative of $C_E(\tsnr)$ with respect to SNR is easily seen to be given by
\begin{gather}
\dot C_{Ei}(\tsnr) = -\frac{1}{\theta_i }\frac{\dot \g_i(\tsnr)}{\g_i(\tsnr)} \label{eq:dotcei0}
\end{gather}
where $\dot{\g}_i(\tsnr)$ denote the first derivative of the function $\g_i(\tsnr)$ with respect to $\tsnr$. It can be readily seen that $\g_i(0)=1$. If we use $f_i(\tsnr)$ as the instantaneous service rate in nats (i.e. $R_i(\tsnr)=f_i(\tsnr)\log_e2$), then we have the relation
\begin{gather}
\dot{\g}_i(0)=-\frac{\theta_i}{\log_e2}\E\left\{\dot{f}_i(0)\right\} \label{eq:dotgi0}
\end{gather}
where the first derivative expressions $\dot{f}_i(0)$ for $i =0,1,2$ are given by
\begin{align} \label{eq:dotfis}
\dot{f}_1(0)&= \delta_1 \left(z_1 - z_2\right) \ind\!\left\{z_1\!\geq \!z_2\right\}, \nonumber
\\
\dot{f}_2(0)&= \delta_2 \left( z_2-z_1 \right) \ind\!\left\{z_1\!< \!z_2\right\}, \nonumber
\\
\dot{f}_0(0)&= (1-\delta_1) z_2\ind\!\left\{z_1\!\geq \!z_2\right\}+ (1-\delta_2) z_1\ind\!\left\{z_1\!< \!z_2\right\}.
\end{align}
By inserting $\dot{f}_i(0)$ formulations above to \eqref{eq:dotgi0}, and then $\dot{\g}_i(0)$ to \eqref{eq:dotcei0} consecutively, we obtain the minimum energy per bit expressions for confidential and common messages in \eqref{eq:EbN01} - \eqref{eq:EbN00} using \eqref{eq:ebnomin_conf2} and \eqref{eq:ebnomin_common2}.

\subsection{Proof of Proposition \ref{prop:ebno_discrete}}  \label{subsec:ebno_discrete}
First, we prove the result for the discrete Markov source. We need to obtain the first derivative of $r_{\avg,i}^*(\tsnr)$. Let us rewrite the maximum average arrival rate in \eqref{eq:2discreteravg} as
\begin{align}
r_{\avg,i}^*(\tsnr,\theta_i)=\frac{P_{\on}}{\theta_i}&\Big[\log_e(1-p_{11}\g_i(\tsnr)) -\log_e(\g_i(\tsnr))\Big. \nonumber
\\
&\Big.-\log_e\big((1-p_{11}-p_{22})\g_i(\tsnr)+p_{22}\big)\Big]
\end{align}
where $\g_i(\tsnr)$ is defined in \eqref{eq:gdef}. Taking the first derivative with respect to $\tsnr$, we obtain
\begin{align}
\dot{r}_{\avg,i}^*(\tsnr,\theta_i)=\frac{P_{\on}}{\theta_i}& \Bigg[\frac{-p_{11}\dot{\g_i}(\tsnr)}{1-p_{11}\g_i(\tsnr)} -\frac{\dot{\g_i}(\tsnr)}{\g_i(\tsnr)}\Big. \nonumber
\\
&\Big.-\frac{(1-p_{11}-p_{22})\dot{\g_i}(\tsnr)}{(1-p_{11}-p_{22})\g_i(\tsnr)+p_{22}}\Bigg]. \label{eq:dotravg_disc}
\end{align}
When we let $\tsnr \to 0$, the first derivative expression becomes
\begin{align}
\dot{r}_{\avg,i}^*(0)&=\frac{\dot{\g}_i(0)}{\theta_i} P_{\on}\Bigg[-\frac{p_{11}}{1-p_{11}} -1-\frac{1-p_{11}-p_{22}}{1-p_{11}} \Bigg]
\\
&=-\frac{\dot \g_i(0)}{\theta}=\frac{\dot{f}_i(0)}{\log_e2}\label{eq:dotravg0_disc}
\end{align}
where $P_{\on}=\tfrac{1-p_{11}}{2-p_{11}-p_{22}}$. Note that $\g(0)=1$. Plugging the result in \eqref{eq:dotfis} and \eqref{eq:dotravg0_disc} into \eqref{eq:ebnomin_conf2} and \eqref{eq:ebnomin_common2}, we immediately obtain (\ref{eq:EbN01}) - (\ref{eq:EbN00}). 

Now, we show the proof for the Markov fluid source. We evaluate the derivative of $r_{\avg,i}^*(\tsnr)$ in \eqref{eq:2fluidravg} with respect to $\tsnr$ and obtain \eqref{eq:dotravg_fluid} given at the top of the next page.
\begin{figure*}
\begin{align}
\dot r_{\avg,i}^*(\tsnr, \theta_i)=-\frac{P_{\on}}{\theta_i} \left\{\log_e\g_i(\tsnr)\tfrac{\operatorname{d}\!}{\operatorname{d}\!\ssnr} \Bigg[\frac{\alpha+\beta-\log_e\g_i(\ssnr)}{\alpha-\log_e\g_i(\ssnr)}\Bigg]+ \frac{\alpha+\beta-\log_e\g_i(\ssnr)}{\alpha-\log_e\g_i(\ssnr)} \frac{\dot{\g_i}(\tsnr)}{\g_i(\tsnr)} \right\} \label{eq:dotravg_fluid}
\end{align}
\end{figure*}
When we let $\tsnr \to 0$, the first derivative expression simplifies to
\begin{align}
\dot{r}_{\avg,i}^*(0)&=-\frac{P_{\on}}{\theta_i} \frac{\alpha+\beta}{\alpha} \dot{\g}_i(0)=\frac{\dot{f}_i(0)}{\log_e2} \label{eq:dotravg0_fluid}
\end{align}
where $P_{\on}=\tfrac{\alpha}{\alpha+\beta}$. Note that $\g(0)=1$. Plugging the result in \eqref{eq:dotfis} and \eqref{eq:dotravg0_fluid} into \eqref{eq:ebnomin_conf2} and \eqref{eq:ebnomin_common2}, we immediately obtain (\ref{eq:EbN01}) - (\ref{eq:EbN00}). 

\subsection{Proof of Proposition \ref{prop:ebno_MMPP}}  \label{subsec:ebno_MMPP}

The proof is straightforward as we note that the maximum average arrival rate $r_{\avg,i}^*(\tsnr)$ of discrete-time MMPP source in \eqref{eq:ravgdMMPP} is the scaled version of that of the discrete Markov source in \eqref{eq:2discreteravg}. The scaling factor is $\frac{\theta_i}{e^\theta_i-1}$. The same assertion can be made for the relationship between the maximum average arrival rates of continuous-time MMPP in \eqref{eq:2MMPPr} and Markov fluid source in \eqref{eq:2fluidravg}. Therefore, the minimum energy per bit expressions for discrete-time and continuous-time MMPP sources can be obtained by scaling the formulations in \eqref{eq:EbN01}-\eqref{eq:EbN00} with $\frac{e^\theta_i-1}{\theta_i}$.

\subsection{Proof of Proposition \ref{prop:widebandslope}} \label{subsec:widebandslope}
Let us recall that the wideband slope is given by
\begin{equation}\label{eq:widebandsloperavg}
\mathcal{S}_0=-\frac{2\big(\dot{r}_\avg^*(0)\big)^2} { \ddot{r}_\avg^*(0)}\log_e{2}.
\end{equation}
When the arrival rate is constant, we can exchange $r_{\avg,i}^*(\tsnr)$ with $C_{Ei}(\tsnr)$. For the wideband slope, in addition to the first derivative of the throughput, we also need to obtain the second derivative of the throughput.
Second derivatives of the effective capacity at $\tsnr=0$ can be computed as
\begin{gather}
\ddot C_{Ei}(\tsnr) = -\frac{1}{\theta_i }\left[\frac{\ddot \g_i(\tsnr)}{\g_i(\tsnr)}-\left(\frac{\dot \g_i(\tsnr)}{\g_i(\tsnr)}\right)^2 \right]. \label{eq:ddotcei0}
\end{gather}
To simplify this equation, we derive the second derivative of $\g_i(\tsnr)$ at $\tsnr=0$ as
\begin{gather}
\ddot{\g}_i(0)=-\frac{\theta_i}{\log_e2} \E\left\{\ddot{f}_i(0)\right\}+ \left(\frac{\theta_i}{\log_e2}\E\left\{\ddot{f}_i(0)\right\}\right)^2, \label{eq:ddotgi0}
\end{gather}
where the second derivative expressions $\ddot{f}_i(0)$ for $i=0,1,2$ are given by
\begin{align} \label{eq:ddotfis}
\ddot{f}_1(0)&= -\delta_1^2 \left[z_1^2 - z_2^2\right] \ind\!\left\{z_1\!\geq \!z_2\right\}, \nonumber
\\
\ddot{f}_2(0)&= -\delta_2^2 \left[ z_2^2-z_1^2 \right]\ind\!\left\{z_1\!< \!z_2\right\}, \nonumber
\\
\ddot{f}_0(0)&= -(1-\delta_1^2)z_2^2\ind\!\left\{z_1\!\geq \!z_2\right\}-(1-\delta_2^2)z_1^2 \ind\!\left\{z_1\!< \!z_2\right\}.
\end{align}

We insert $\dot{f}_i(0)$ in \eqref{eq:dotfis} and $\ddot{f}_i(0)$ in \eqref{eq:ddotfis} onto $\dot{\g}_i(0)$ in \eqref{eq:dotgi0} and $\ddot{\g}_i(0)$ in \eqref{eq:ddotgi0} in order to obtain first and second derivative expressions of the effective capacity. By incorporating the \eqref{eq:dotcei0} and \eqref{eq:ddotcei0} on \eqref{eq:widebandsloperavg} we obtain wideband slope expression in \eqref{eq:S0i}.

\subsection{Proof of Proposition \ref{prop:wbsdisc}} \label{subsec:wbsdisc}

In order to find the wideband slope, we need to determine the second derivative of the maximum average arrival rate with respect to $\tsnr$. As $\tsnr\rightarrow0$ the second derivative expression is given by
\begin{align}
\ddot{r}_{\avg,i}^*(0,\theta_i)=&\frac{\ddot{\g_i}(0)}{\theta_i} P_{\on}\Bigg[-\frac{p_{11}}{1-p_{11}} -1-\frac{(1-p_{11}-p_{22})}{1-p_{11}} \Bigg] \nonumber
\\
&+\frac{\left[\dot{\g_i}(0)\right]^2}{\theta_i}P_{\on}\Bigg[-\frac{p_{11}^2}{(1-p_{11})^2} +1\!+\frac{(1-p_{11}-p_{22})^2}{(1-p_{11})^2} \Bigg]\nonumber
\\
=&-\frac{\ddot{\g_i}(0)}{\theta_i}+(1-\eta)\frac{\left[\dot{\g_i}(0)\right]^2}{\theta_i} \label{eq:ddotravg0_disc}
\end{align}
where $\eta$ is defined in (\ref{eq:eta}). The fact that $g_i(0) = 1$ is taken into account in (\ref{eq:ddotravg0_disc}). Finally, inserting \eqref{eq:dotravg0_disc} and \eqref{eq:ddotravg0_disc} into \eqref{eq:widebandsloperavg}, the wideband slope expression in \eqref{eq:S0idisc} is readily obtained.

\subsection{Proof of Proposition \ref{prop:wbsfluid}} \label{subsec:wbsfluid}

In order to find the wideband slope, we need to determine the second derivative of the maximum average arrival rate with respect to $\tsnr$. When $\tsnr\rightarrow0$, the second derivative expression is given by
\begin{align}
\ddot{r}_{\avg,i}^*(0,\theta_i)=&-\frac{P_{\on}}{\theta_i} \left\{\frac{2\beta}{\alpha^2}\dot{\g_i}(0)+\frac{\alpha+\beta}{\alpha}\left(\ddot{\g_i}(0)-\dot{\g_i}(0)\right)\right\}\nonumber
\\
=&-\frac{\ddot{\g_i}(0)}{\theta_i}+(1-\zeta)\frac{\left[\dot{\g_i}(0)\right]^2}{\theta_i} \label{eq:ddotravg0_fluid}
\end{align}
where $\zeta$ is defined in (\ref{eq:zeta}) and we again use the fact that $\g_i(0) = 1$. Finally, inserting \eqref{eq:dotravg0_fluid} and \eqref{eq:ddotravg0_fluid} into \eqref{eq:widebandsloperavg}, we obtain the wideband slope expression in \eqref{eq:S0ifluid}.

\subsection{Proof of Proposition \ref{prop:impcsi}} \label{subsec:impcsi}

First, we define
\begin{equation}
\g(\tsnr)=1-\p\{\Gamma_1\}\left(1-e^{-\theta \lambda}\right). \label{ravgdiscV2}
\end{equation}
The maximum average arrival rate of the ON-OFF discrete Markov source can be rewritten as
\begin{align}
r_\avg^*(\tsnr)=\frac{P_{\on}}{\theta}\left[\log_e\left(\frac{1-p_{11}\g(\tsnr)}{(1\!-\!p_{11}\!-\!p_{22})\g^2(\tsnr)+p_{22}\g(\tsnr)}\right)\right].
\end{align}

In order to find the minimum energy per bit and wideband slope, we need to determine the first and second derivatives of the maximum average arrival rate with respect to $\tsnr$. Initially, we take the first derivative of maximum average arrival rate and let $\tsnr \to 0$ as follows:
\begin{align}
\dot{r}_\avg^*(0)&=\frac{\dot{\g}(0)}{\theta} P_{\on}\Bigg[-\frac{p_{11}}{1-p_{11}} -1-\frac{1-p_{11}-p_{22}}{1-p_{11}} \Bigg]
\\
&=-\frac{\dot \g(0)}{\theta}. \label{eq:dotravg0_disc2}
\end{align}
For this, we also need to characterize the first derivative of $\g(\tsnr)$. We start with the Taylor series expansion of the fixed rate $\lambda$ in the low-SNR regime:
\begin{equation}
\lambda=\frac{a}{\log_e2}\tsnr + \frac{b}{\log_e2}\tsnr^2 + o(\tsnr^2) \label{eq:lambdataylor}.
\end{equation}
Now, the first derivative of $\g(\tsnr)$ is given by
\begin{equation} \label{eq:dotgsnrimpcsi}
\dot{\g}(\tsnr)=-\frac{\partial}{\partial\tsnr}\p\{\Gamma_1\}\left(1-e^{-\theta \lambda}\right)+\p\{\Gamma_1\}\frac{\partial e^{-\theta \lambda}}{\partial\tsnr}.
\end{equation}
As $\tsnr\rightarrow0$, we have $\lambda\rightarrow0$. Therefore at $\tsnr=0$, we have
\begin{equation}\label{eq:dotg0impcsi}
\dot{\g}(0)=\lim_{\tsnr\to0}\p\{\Gamma_1\} (-\theta)e^{-\theta \lambda} \frac{\partial \lambda}{\partial\tsnr}.
\end{equation}

To proceed we need to obtain the probability expression $\p\{\Gamma_1\}$. For independent and exponentially distributed $z_1$ and $z_2$ with unit mean, we can obtain
\begin{align}
\p\{\Gamma_1\} &= \int_0^\infty e^{-z_2}\int_{2^\lambda z_2+ \frac{2^\lambda-1}{\ssnr}}^\infty e^{-z_1}dz_1 dz_2
\\
&=e^{-\frac{2^\lambda-1}{\ssnr}}\int_0^\infty e^{-(2^\lambda+1)z_2} dz_2
\\
&=e^{-\frac{2^\lambda-1}{\ssnr}}\frac{1}{2^\lambda+1}.
\end{align}
Now, we can simplify the expression in \eqref{eq:dotg0impcsi} as
\begin{equation}
\dot{\g}(0)=-\frac{e^{-a}}{2}\theta \frac{a}{\log_e2}, \label{eq:dotg0impcsiv2}
\end{equation}
and inserting this expression into \eqref{eq:ebnomin-ra}, we obtain the minimum energy per bit as
\begin{equation}
\frac{E_b}{N_0}_{\tmin}=-\frac{\theta}{\dot{\g}(0)}=\frac{2\log_e2}{ae^{-a}}.
\end{equation}
Finally, we want to determine the smallest possible minimum energy per bit expression. It can be easily seen that the smallest value for the minimum energy per bit is obtained when $a=1$, leading to the minimum energy per bit expression in \eqref{eq:impcsiebno}.

In order to find the wideband slope, we first determine the second derivative of the maximum average arrival rate with respect to $\tsnr$ and then evaluate it at $\tsnr = 0$. The resulting equation is given above in \eqref{eq:ddotravg0_disc2}.
\begin{figure*}
\begin{align}
\ddot{r}_\avg^*(0)=&\frac{\ddot{\g}(0)}{\theta} P_{\on}\Bigg[-\frac{p_{11}}{1-p_{11}} -1-\frac{(1-p_{11}-p_{22})}{1-p_{11}} \Bigg] +\frac{\left[\dot{\g}(0)\right]^2}{\theta}P_{\on}\Bigg[-\frac{p_{11}^2}{(1-p_{11})^2} +1\!+\frac{(1-p_{11}-p_{22})^2}{(1-p_{11})^2} \Bigg]\nonumber
\\
=&-\frac{\ddot{\g}(0)}{\theta}+(1-\eta)\frac{\left[\dot{\g}(0)\right]^2}{\theta} \label{eq:ddotravg0_disc2}.
\end{align}
\end{figure*}
Note that, $\eta$ is defined in \eqref{eq:eta}. The first derivative of $\g(\tsnr)$ at $\tsnr=0$ is given by \eqref{eq:dotg0impcsi}, and the second derivative is
\begin{align}
\ddot{\g}(0)=&\lim_{\tsnr\to0}2\frac{\partial \p\{\Gamma_1\}}{\partial\tsnr}\frac{\partial e^{-\theta \lambda}}{\partial\tsnr}+\p\{\Gamma_1\}\frac{\partial e^{-\theta \lambda}}{\partial\tsnr^2}
\\
=&2\left(-\frac{1}{4}e^{-a}(a^2+a+2b)(-\theta)\frac{a}{\log_e2}\right)\nonumber
\\
&+\frac{e^{-a}}{2}\left(-\theta\frac{2b}{\log_e2}+\theta^2\frac{a^2}{(\log_e2)^2}\right)
\\
=&\frac{e^-a}{2} \left[\theta \frac{a^3+a^2}{\log_e2}+\theta^2 \frac{a^2}{(\log_e2)^2}+\theta \frac{2b}{\log_e2}(a-1) \right]. \label{eq:ddotg0impcsi}
\end{align}

The wideband slope expression can be determined inserting the first and second derivative expressions in \eqref{eq:dotravg0_disc2} and \eqref{eq:ddotravg0_disc2} into \eqref{eq:widebandslope-ra}:
\begin{align}
\mathcal{S}_{0}&=\frac{2 \frac{\left(\dot{\g}(0)\right)^2}{\theta^2} }{ \frac{\ddot{\g}(0) }{\theta}+\frac{\eta-1}{\theta} \left(\dot{\g}(0)\right)^2} \log_e2
\\
&=\frac{1}{\frac{\theta (\eta-1)}{2\log_e2}+\frac{\theta}{\log_e2 e^{-a}}+\frac{a+1}{e^{-a}}+\frac{2b(a-1)}{a^2 e^{-a}} }. \label{eq:widebandslopeinab}
\end{align}
Since the wideband slope is defined as the slope at the minimum energy per bit, we set $a=1$. Note that with this choice, parameter $b$ vanishes as $2b(a-1)\to 0$ in \eqref{eq:widebandslopeinab}. Thus, we obtain the formulation in \eqref{eq:impcsiS0}.

\subsection{Proof of Proposition \ref{prop:impcsifluid}} \label{subsec:impcsifluid}

The maximum average arrival rate of Markov fluid source can be rewritten as
\begin{align}
r_\avg^*(\tsnr)=-\frac{P_{\on}}{\theta}\left[1+\frac{\beta}{\alpha-\log_e(\g(\tsnr))}\right]\log_e(\g(\tsnr)). \label{eq:ravgfluidv2}
\end{align}

By taking the first derivative of the expression in \eqref{eq:ravgfluidv2} and letting $\tsnr \to 0$, we obtain the following:
\begin{align}
\dot{r}_\avg^*(0)&=-\frac{P_{\on}}{\theta}\left[1+\frac{\beta}{\alpha}\right]\dot{\g}(0)=-\frac{\dot \g(0)}{\theta}. \label{eq:dotravg0_fluid2}
\end{align}

By combining \eqref{eq:dotravg0_fluid2} with \eqref{eq:dotg0impcsiv2} as $a\to 1$, and inserting into \eqref{eq:ebnomin-ra}, we obtain the minimum energy per bit given in \eqref{eq:impcsiebnofluid}.

Next, we take the second derivative of the maximum average arrival rate with respect to $\tsnr$ and then evaluate it at $\tsnr = 0$ as
\begin{align}
\ddot{r}_\avg^*(0)=&-\frac{P_{\on}}{\theta}\left[\left(\frac{2\beta}{\alpha^2}-1-\frac{\beta}{\alpha}\right)\left(\dot{\g}(0)\right)^2 + \left(1+\frac{\beta}{\alpha}\right) \ddot{\g}(0)  \right]
\\
=&-\frac{\ddot{\g}(0)}{\theta}+(1-\zeta)\frac{\left[\dot{\g}(0)\right]^2}{\theta} \label{eq:ddotravg0_fluid2}.
\end{align}
Note that, $\zeta$ is defined in \eqref{eq:zeta}. We derive the wideband slope expression by using \eqref{eq:dotravg0_fluid2}, \eqref{eq:ddotravg0_fluid2} and \eqref{eq:widebandslope-ra}. Again, since the wideband slope is defined at the minimum energy per bit, we set $a=1$ and obtain the formulation in \eqref{eq:impcsiS0fluid}.

\end{document}